%% file: main.tex
\documentclass[twocolumn,twocolappendix]{aastex7}

\usepackage{threeparttable}
\usepackage{color}
\usepackage{amsmath}
\usepackage{ulem}

\input{extra_preamble.tex}

\definecolor{ForestGreen}{RGB}{34,139,34}
\def\tw#1 {{\textcolor{ForestGreen}{#1}}\ }
\def\blue#1 {{\textcolor{blue}{#1}}\ }
\def\red#1 {{\textcolor{red}{#1}}\ }
\begin{document} 
\title{JWST/MIRI reveals the true number density of massive galaxies in the early Universe}

\author[0000-0002-2504-2421,sname=Wang,gname=Tao]{Tao Wang}
\affiliation{School of Astronomy and Space Science, Nanjing University, Nanjing 210093, China}
\affiliation{Key Laboratory of Modern Astronomy and Astrophysics, Nanjing University, Ministry of Education, Nanjing 210093, China}
\email[show]{taowang@nju.edu.cn}

\author{Hanwen Sun}
\affiliation{School of Astronomy and Space Science, Nanjing University, Nanjing 210093, China}
\affiliation{Key Laboratory of Modern Astronomy and Astrophysics, Nanjing University, Ministry of Education, Nanjing 210093, China}
\email{hanwensun@smail.nju.edu.cn}
\author{Luwenjia Zhou}
\affiliation{School of Astronomy and Space Science, Nanjing University, Nanjing 210093, China}
\affiliation{Key Laboratory of Modern Astronomy and Astrophysics, Nanjing University, Ministry of Education, Nanjing 210093, China}
\email{wenjia@nju.edu.cn}

\author{Ke Xu}
\affiliation{School of Astronomy and Space Science, Nanjing University, Nanjing 210093, China}
\affiliation{Key Laboratory of Modern Astronomy and Astrophysics, Nanjing University, Ministry of Education, Nanjing 210093, China}
\email{MG1926012@smail.nju.edu.cn}

\author{Cheng Cheng}
\affiliation{Chinese Academy of Sciences South America Center for Astronomy, National Astronomical Observatories, CAS, Beijing, 100101, China}
\email{chengcheng@nao.cas.cn}

\author{Zhaozhou Li}
\affiliation{Center for Astrophysics and Planetary Science, Racah Institute of Physics, The Hebrew University, Jerusalem, 91904, Israel}
\email{zhaozhou.li@mail.huji.ac.il}

\author{Yangyao Chen}
\affiliation{School of Astronomy and Space Science, University of Science and Technology of China, Hefei, Anhui 230026, China}
\affiliation{Key Laboratory for Research in Galaxies and Cosmology, Department of Astronomy, University of Science and Technology of China, Hefei, Anhui 230026, China}
\email{yangyaochen.astro@foxmail.com}

\author{H. J. Mo}
\affiliation{Department of Astronomy, University of Massachusetts, Amherst, MA 01003-9305, USA}
\email{hjmo@umass.edu}

\author{Avishai Dekel}
\affiliation{Center for Astrophysics and Planetary Science, Racah Institute of Physics, The Hebrew University, Jerusalem, 91904, Israel}
\affiliation{Santa Cruz Institute for Particle Physics, University of California, Santa Cruz, CA 95064, USA}
\email{avishai.dekel@mail.huji.ac.il}

\author{Tiancheng Yang}
\affiliation{School of Astronomy and Space Science, Nanjing University, Nanjing 210093, China}
\affiliation{Key Laboratory of Modern Astronomy and Astrophysics, Nanjing University, Ministry of Education, Nanjing 210093, China}
\email{652023260012@smail.nju.edu.cn}

\author{Yijun Wang}
\affiliation{School of Astronomy and Space Science, Nanjing University, Nanjing 210093, China}
\affiliation{Key Laboratory of Modern Astronomy and Astrophysics, Nanjing University, Ministry of Education, Nanjing 210093, China}
\email{wangyijun@nju.edu.cn}

\author{Longyue Chen}
\affiliation{School of Astronomy and Space Science, Nanjing University, Nanjing 210093, China}
\affiliation{Key Laboratory of Modern Astronomy and Astrophysics, Nanjing University, Ministry of Education, Nanjing 210093, China}
\email{652023260002@smail.nju.edu.cn}

\author{Xianzhong Zheng}
\affiliation{Purple Mountain Observatory, Chinese Academy of Sciences, 10 Yuanhua Road, Qixia District, Nanjing 210023, China}
\affiliation{School of Astronomy and Space Science, University of Science and Technology of China, Hefei, Anhui 230026, China}
\email{xzzheng@pmo.ac.cn}

\author{Zheng Cai}
\affiliation{Department of Astronomy, Tsinghua University, Beijing 100084, China}
\email{zcai@mail.tsinghua.edu.cn}

\author{David Elbaz}
\affiliation{AIM, CEA, CNRS, Université Paris-Saclay, Université Paris
Diderot, Sorbonne Paris Cité, F-91191 Gif-sur-Yvette, France}
\email{delbaz@cea.fr}

\author{Y.-S. Dai}
\affiliation{Chinese Academy of Sciences South America Center for Astronomy, National Astronomical Observatories, CAS, Beijing, 100101, China}
\email{ydai@nao.cas.cn}

\author{J.-S. Huang}
\affiliation{Chinese Academy of Sciences South America Center for Astronomy, National Astronomical Observatories, CAS, Beijing, 100101, China}
\email{jhuang@nao.cas.cn}
   
\begin{abstract}
Early JWST studies reporting an unexpected abundance of massive galaxies at $z \sim 5$--$8$ challenge galaxy formation models in the $\Lambda$CDM framework. Previous stellar mass ($M_\star$) estimates suffered from large uncertainties due to the lack of rest-frame near-infrared data. Using deep JWST/NIRCam and MIRI photometry from PRIMER, we systematically analyze massive galaxies at $z \sim 3$--$8$, leveraging rest-frame $\gtrsim 1\,\mu$m constraints. We find MIRI is critical for robust $M_\star$ measurements for massive galaxies at $z > 5$: excluding MIRI overestimates $M_\star$ by $\sim 0.4$ dex on average for $M_\star > 10^{10}\,M_\odot$ galaxies, with no significant effects at lower masses. This reduces number densities of $M_\star > 10^{10}\,M_\odot$ ($10^{10.3}\,M_\odot$) galaxies by $\sim 36\%$ ($55\%$). MIRI inclusion also reduces ``Little Red Dot'' (LRD) contamination in massive galaxy samples, lowering the LRD fraction from $\sim 32\%$ to $\sim 13\%$ at $M_\star > 10^{10.3}\,M_\odot$. Assuming pure stellar origins, LRDs exhibit $M_\star \sim 10^{9\text{--}10.5}\,M_\odot$ with MIRI constraints, rarely exceeding $10^{10.5}\,M_\odot$. Within standard $\Lambda$CDM, our results indicate a moderate increase in the baryon-to-star conversion efficiency ($\epsilon$) toward higher redshifts and masses at $z > 3$. For the most massive $z \sim 8$ galaxies, $\epsilon \sim 0.3$, compared to $\epsilon \lesssim 0.2$ for typical galaxies at $z < 3$. This result is consistent with models where high gas densities and short free-fall times suppress stellar feedback in massive high-$z$ halos.

\end{abstract}


\section{introduction}
\label{sec:intro}

In the \lcdm\ cosmological model, the most massive galaxies tend to populate the most massive dark matter halos. Consequently, the number density of massive galaxies provides crucial constraints on galaxy formation models and cosmology. Over the last decade, our understanding of massive galaxies ($\mstar \gtrsim 10^{10}$ M$_{\odot}$) in the early Universe has improved significantly. A substantial population of massive, ultraviolet (UV)-faint galaxies at $z \gtrsim 4$ have been confirmed by near-infrared (NIR) to millimeter spectroscopy~\citep{HuangJ:2011,Walter:2012,Franco:2018,WangT:2019,Yamaguchi:2019}, which were absent from previous studies focusing on UV-bright galaxies selected via the Lyman-break technique. Most of them are likely dusty and star-forming, as indicated by high submillimeter detection rates~\citep{WangT:2019}. A subset, however, may represent the  first population of quiescent galaxies, where red optical-to-NIR colors arise from evolved stellar populations.
{\it JWST} observations have further confirmed the existence of these massive galaxies and extended it to higher redshifts, including both dusty and quiescent galaxies~\citep{Labbe:2023,Barro:2024,WangB_Rubies:2024,weibel_rubies_2024,glazebrook_massive_2024,carnall_massive_2023,kakimoto_massive_2024}.

The new {\it JWST} observations have raised new challenges regarding both the high number density of massive galaxies and their stellar masses at $z \sim 5-9$~\citep{Boylan-Kolchin:2023,XiaoM:2023}. A similar phenomenon exists at $z > 10$ for luminous UV-selected galaxies~\citep{YanH:2023,Harikane:2023a,McLeod:2024}.
If confirmed, this would require either significantly elevated baryon conversion efficiency that is unexpected by current galaxy formation models, or a larger number of massive dark matter (DM) halos exceeding \lcdm\ predictions. Recent studies show that many massive galaxies selected in early {\it JWST} studies are Type-I AGN candidates with pointlike morphologies \citep[``Little Red Dots" (LRDs); e.g., ][]{Matthee:2024,Greene:2024,Kocevski:2025,Baggen:2024LRD,PerezGonzalez:2024}, for which stellar mass estimates are highly uncertain. Nevertheless, not all massive galaxies in the early Universe are LRDs. Determining the true number density of massive galaxies with reliable $\mstar$ estimates represents one of the most urgent tasks in extragalactic astronomy.

Previous studies of massive galaxies in the early Universe suffer from two major limitations: small number statistics and large uncertainties in $\mstar$ estimation due to the lack of rest-frame NIR photometry. Recent studies indicate that without {\it JWST} Mid-Infrared Instrument (MIRI) photometry probing rest-frame $ \gtrsim 1 \mu$m, $\mstar$ can be overestimated by $\sim 0.4$ dex at $z \sim 5-9$~\citep{Papovich:2023,SongJ:2023,Williams:2024,WangB:2024a}.
These results suggest that a census of high-redshift massive galaxies using mid-infrared 
photometry over large volumes is key to confirming or falsifying tensions between observations and current galaxy formation models in the \lcdm\ cosmology. 

In this study, we use data from one of the largest and deepest {\it JWST}/NIRCam and MIRI surveys--the Public Release IMaging for Extragalactic Research (PRIMER, GO 1837, PI: James Dunlop,~\citealt{Dunlop2021}) program--to systematically study massive galaxies at $3 < z < 8$. The unprecedented multiwavelength data and large area allow stringent constraints on the number density and stellar mass functions (SMFs) of massive galaxies in the early Universe. The Letter is structured as follows: Section~\ref{Sec:data} presents the data, sample construction, and derivation of physical parameters. Section~\ref{sec:results} presents the main results, including differences in massive galaxy selection with and without (w/wo) MIRI photometry, the derived SMFs, and implications for cosmic evolution of baryon conversion efficiency. Section~\ref{sec:discussion} discusses the results, and Section~\ref{sec:conclusion} summarizes our conclusions.

Throughout this paper, we adopt a cosmological model with ${H_0}{\rm{ = }}70~{\rm{km}} \cdot {{\rm{s}}^{ - 1}} \cdot {\rm{Mp}}{{\rm{c}}^{ - 1}}$, ${\Omega _{\rm{m}}} = 0.3$ and ${\Omega _\Lambda } = 0.7$. Magnitudes use the AB system \citep{Oke1983}, and stellar masses are estimated assuming a \citet{Kroupa2001} initial mass function (IMF).

\section{Data and Sample construction }\label{Sec:data}

\subsection{PRIMER survey and multiwavelength data}
The PRIMER survey targets the CANDELS-COSMOS and CANDELS-UDS fields with JWST/NIRcam and MIRI. Deep imaging was obtained in 10 bands: F090W, F115W, F150W, F200W, F277W, F356W, F444W and F410M with NIRCam; and F770W and F1800W with MIRI. Its survey volume is $\sim 8$ times larger than that of \citet{Labbe:2023}, and deep MIRI/F770W and F1800W coverage enable a robust census of massive galaxies in the early Universe. In addition to PRIMER, we include available {\it JWST}/NIRCam and MIRI imaging data from other surveys covering (parts of) the PRIMER fields, COSMOS-Web~\citep{Casey2023}, PANORAMIC~\citep{Williams:2025}, {\it JWST} projects GO 3990~\citep{Morishita2023}, and GO 1840~\citep{Alvarez-Marquez2021}.
The effective areas with NIRCam(F444W)/MIRI(F770W) coverage is 283.2/194.3 arcmin$^{2}$ in CANDELS-COSMOS and 287.9/128.7 arcmin$^{2}$ in CANDELS-UDS. 

We reduced the {\it JWST}/NIRCam and MIRI data using a custom pipeline based on the JWST Calibration Pipeline v1.13.4 \citep{Bushouse2024}, yielding images with improved qualities compared to the default pipeline, especially for MIRI (Appendix~\ref{appendix:MIRI}, Figure~\ref{efig:MIRI-images}).

\begin{figure}
	\centering
     \includegraphics[scale=0.8]{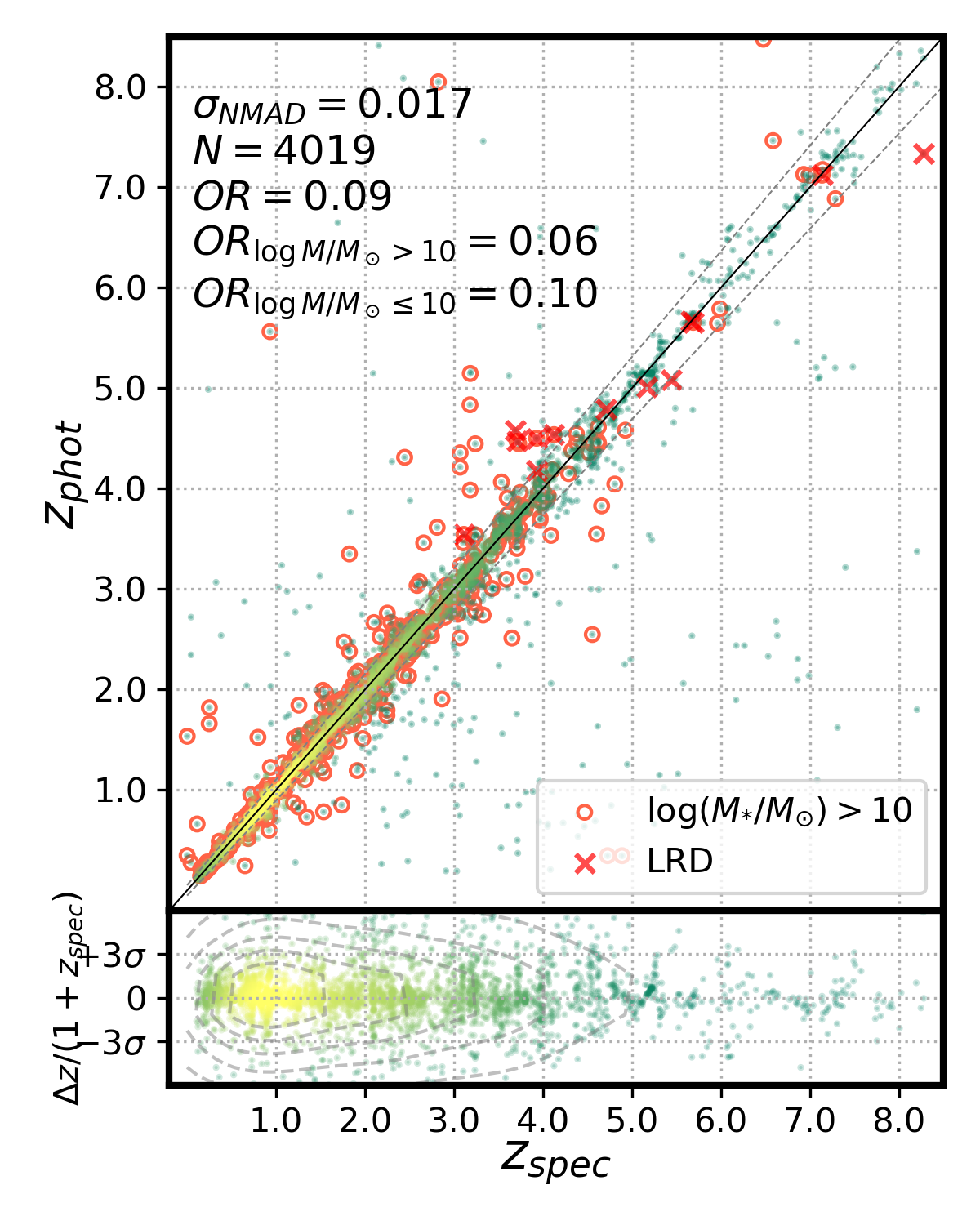}
	\caption{\small{\textbf{Comparison of photometric redshifts ($z_{\rm phot}$) derived with JWST/MIRI photometry against spectroscopic redshifts ($z_{\rm spec}$) for PRIMER galaxies.}
High-mass galaxies with $\mstar > 10^{10} \msun$ are further denoted as red open circles, while LRDs are denoted as red crosses. Outlier rates (QRs, $|\Delta z| > 3\sigma_{\rm NMAD}\times(1+z_{spec})$) for the whole sample and low- and high-mass subsamples are indicated in the top-left corner, which are derived assuming the $\sigma_{\rm NMAD}$ value for the whole sample (0.017). 
$\sigma_{\rm NMAD}$ and ORs show no significant differences between high- and low-mass galaxies.
}}
	\label{fig:pz-specz-comp}
\end{figure}

\begin{figure*}
	\centering
 \includegraphics[width=0.8\textwidth]{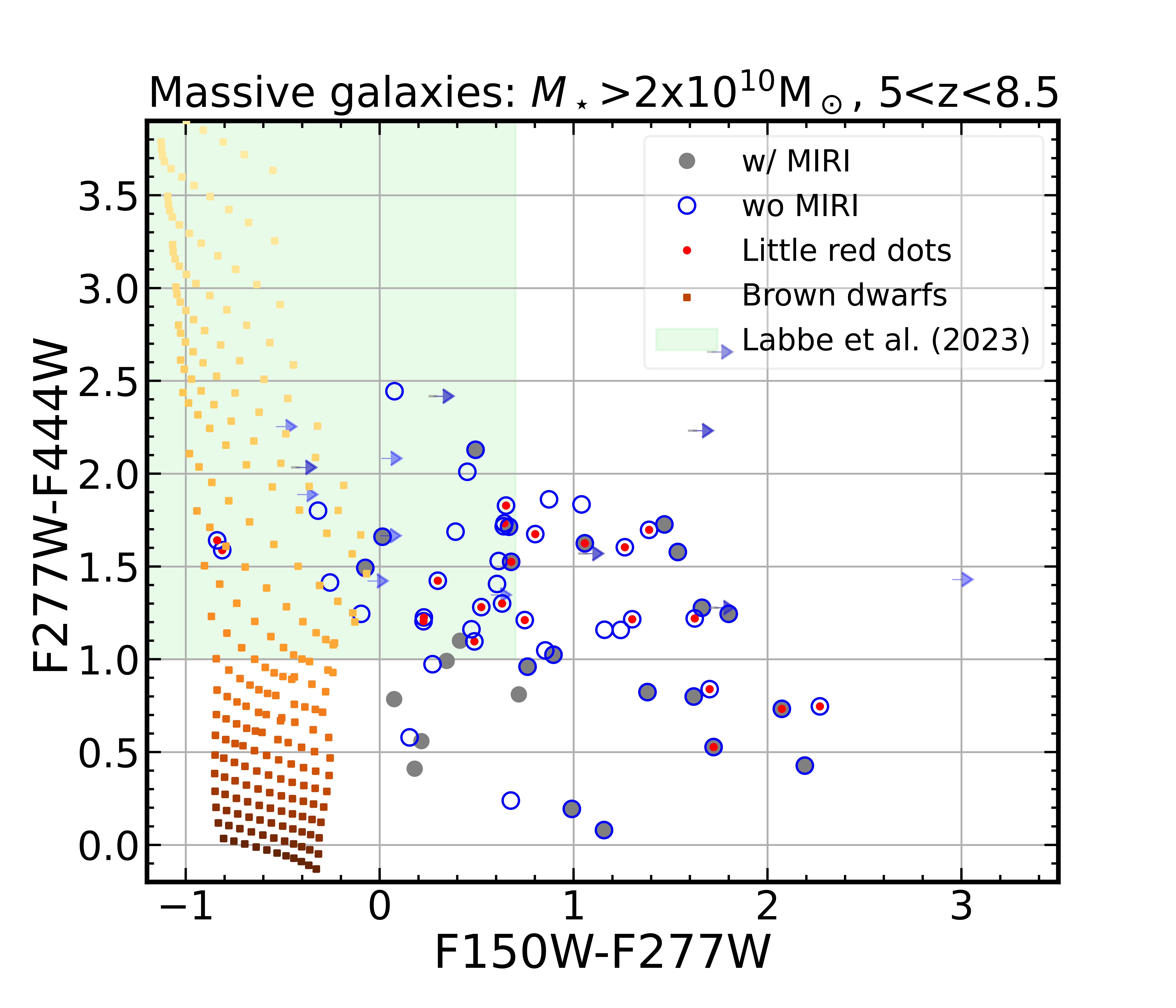}
	\caption{\small{\textbf{Color-color distributions of massive galaxies at $5 < z < 8.5$ identified based on SED-fitting w/wo MIRI photometry.} 
 Only galaxies with $M_{\star} > 2 \times 10^{10} M_{\odot}$ are shown, including sources with either $M_{\star, \rm MIRI} > 2 \times 10^{10} M_{\odot}$ (grey filled circles) or $M_{\star,\rm noMIRI} > 2 \times 10^{10} M_{\odot}$ (blue open circles).  Right-pointing triangles indicate sources undetected in F150W ($S/N < 2$), whose F150W$-$F277W colors are derived from 3$\sigma$ upper limits at F150W.  
 The green shaded region shows the color selection criterion used in~\cite{Labbe:2023} for selecting  $z \gtrsim 6$ massive galaxies with V-shaped SEDs. The colored squares denote the evolution tracks for brown dwarfs~\citep{Langeroodi:2023,Greene:2024} with a set of metallicities ([M/H] = $-0.5$ to $+ 0.5$), C/O ratios (from 0.5 to 1.5 times that of solar), and ages with cloudless atmospheres~\citep{Marley:2021}. The red filled circles denote LRDs. The number of $M_{\star} > 2 \times 10^{10} M_{\odot}$ galaxies drops significantly when MIRI is included, 31 with $M_{\star, \rm MIRI} > 10^{10.3} M_{\odot}$ vs. 68 with $M_{\star, \rm noMIRI} > 10^{10.3} M_{\odot}$. Most MIRI-confirmed massive galaxies lie outside the green shaded region, and are not LRDs (LRD fraction: 4/31 with MIRI vs. 22/68 without MIRI).
 }}
	\label{fig:color_color}
\end{figure*}

We also include archival imaging from $U$-band to 8$\mu$m from space-based ({\it HST} and {\it Spitzer}) and ground-based telescopes (Subaru, CFHT, VISTA, and UKIRT) to extend multiwavelength coverage of galaxy spectral energy distributions (SEDs) in the PRIMER fields. Most galaxies in our sample have photometric measurements across $>30$ bands. Source detection was performed on combined NIRCam long-wavelength (LW) filter images (F277W+F356W+F410M+F444W) using source-extractor v2.25.0 \citep{Bertin1996}. 
Aperture photometry in six apertures (0.2$\arcsec$, 0.3$\arcsec$, 0.4$\arcsec$, 0.5$\arcsec$, 0.7$\arcsec$, 1.0$\arcsec$) and elliptical Kron apertures~\citep{Kron:1980}) using APHOT \citep{Merlin:2019} was  performed on PSF-homogenized images for all {\it HST} and {\it JWST}/NIRCam bands, with their PSFs matched to F444W. Fluxes were corrected to total fluxes via aperture corrections, with the optimal aperture chosen adaptively per source.
For lower-resolution bands (ground-based and {\it JWST}/MIRI), we derived template-fitting deblending photometry to obtain total fluxes using TPHOT v2.0 \citep{Merlin2016}.
Details on archival data and catalog construction will be presented in a forthcoming paper (Sun, H. et al., in prep).
Here in Appendix~\ref{appendix:phot_comp} and Figure~\ref{efig:phot_compare} we compare the photometry in two example bands with other teams, which show excellent agreement.

The final photometric catalog includes 92215 F444W-detected sources with S/N$_{\rm F444W} > 7$ and MIRI/F770W coverage. Of these, 43111 sources are detected in MIRI/F770W with S/N$_{\rm F770W} >$ 2 and the remaining ones have upper limits.

\subsection{Photometric redshifts estimation} 
\label{subsec:photo-zs}

We performed multiple suites of photometric redshift (photo$-z$) and SED fits, with and without MIRI photometry.
Photo-zs were derived using EAZY~\citep{Brammer:2008} with the sfhz\_blue\_13 template set, which includes templates of emission-line dominated sources. A 5\% systematic uncertainty floor was added to the photometric fluxes. 

We explored the influence of MIRI photometry on photo-z estimation by comparing results w/wo MIRI (Appendix~\ref{appendix:photz_comp}). No systematic differences were found with median $z_{\rm noMIRI} - z_{\rm MIRI} \sim 0$ (Figure~\ref{efig:z_comp}). However, many galaxies with $z_{\rm noMIRI} \lesssim 4$ have higher photo-z when MIRI is included.
This likely arises from difficulties in distinguishing optical emission lines from the 1.6$\mu$m stellar bump without MIRI (Figure~\ref{efig:z_comp_example}). Since solutions with MIRI are closer to spectroscopic redshifts (spec-zs), we fix redshifts to the maximum-likelihood values from EAZY with MIRI for SED fitting.

The exquisite multiwavelength data enable highly accurate photo-zs, as validated against spec-zs compiled from the Dawn JWST Archive (DJA) in Figure~\ref{fig:pz-specz-comp}. Most spec-zs for galaxies at $z > 3$ are obtained by the RUBIES survey~\citep{deGraaff:2024}. We characterize the photo-z accuracy by the normalized median absolute deviation ($\sigma_{\rm NMAD}$, \cite{Brammer:2008}). We obtain $\sigma_{\rm NMAD} \sim 0.017$ for the whole sample, with $\sigma_{\rm NMAD} \sim 0.02$ and $\sigma_{\rm NMAD} \sim 0.016$ for high- and low-mass galaxies, respectively. The  outlier rate ($|\Delta z| > 3\sigma_{\rm NMAD}\times(1+z_{spec})$) is $<$~1\%, with no significant difference between low- and high-mass galaxies. 

\begin{figure*}
	\centering
     \includegraphics[scale=0.4]{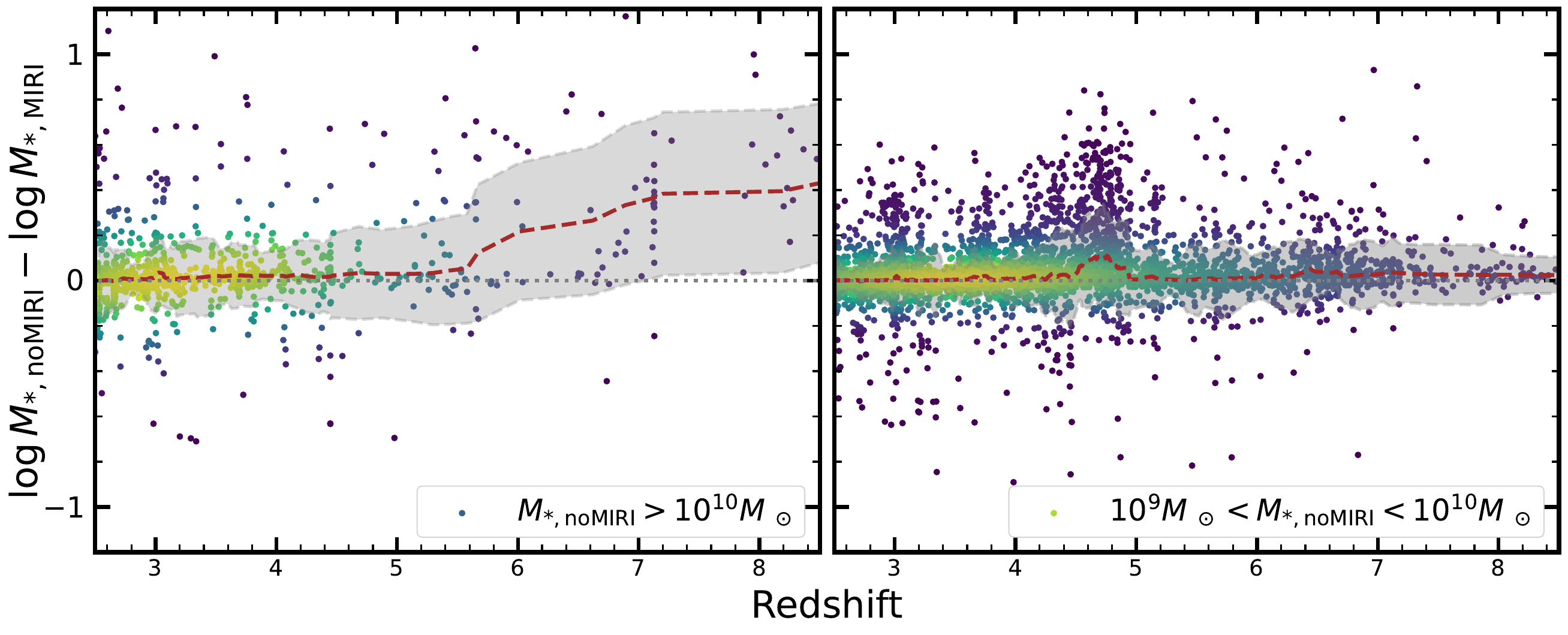}
	\caption{\small{\textbf{Differences in $\mstar$ estimates with and without MIRI vs. redshift for galaxies at $2.5 < z < 8.5$.}
Both panels show only MIRI/F770W-detected galaxies, with the left panel showing the high-mass subsample ($M_{\star,\rm noMIRI} > 10^{10}M_{\odot}$) and the right panel showing the low-mass ones (($M_{\star,\rm noMIRI} < 10^{10}M_{\odot}$). The data points are color-coded by their local density.
The y-axis shows the difference in $\mstar$ estimate w/wo MIRI, which is defined as $\Delta \mstar = {\rm log} (M_{\star, \rm noMIRI}/M_{\star, \rm MIRI})$. 
  The red dashed line indicates the sliding median while the gray shaded region shows the $1\sigma$ (16 to 84 percent) percentiles.
 }}
	\label{fig:mass_diff}
\end{figure*}

\begin{figure*}
	\centering
\includegraphics[width=0.9\textwidth]{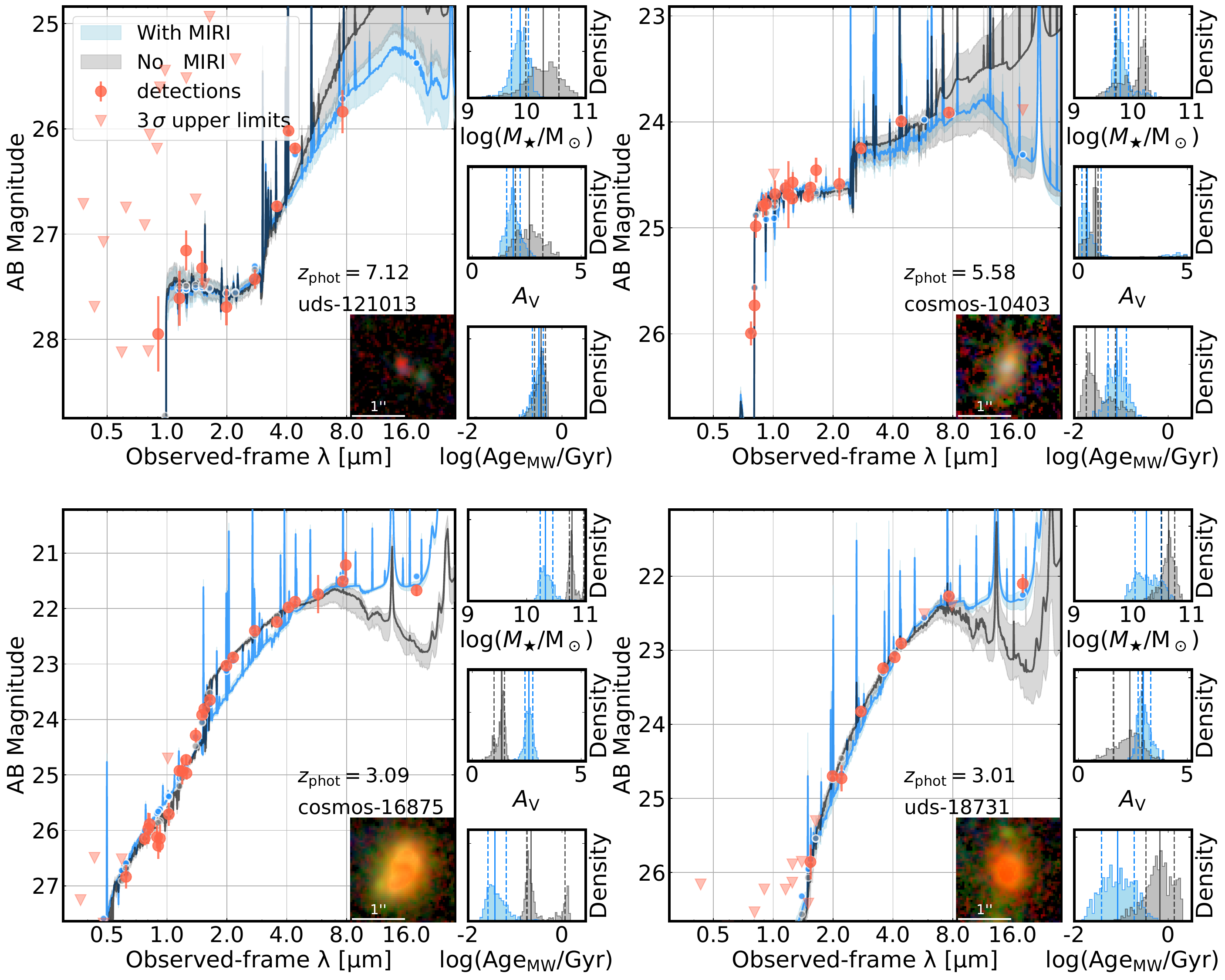}
	\caption{\small{\textbf{Examples SED fits with/without MIRI.} Top: Galaxies at $z > 5$ showing systematic $\mstar$ differences w/wo MIRI. Bottom: Dusty galaxies at $z \sim 3$ with significant $\mstar$ overestimation without MIRI. Orange circles and triangles show measured flux densities or 3$\sigma$ upper limits (for non-detections) in different bands. Blue and black lines show the best-fit SED templates w/wo MIRI bands. The small blue and black points show model photometry. 
 RGB false-color images (R:F444W, G:F277W, B:F150W) are shown in the lower right corner of the corresponding SED plot.
 Adjacent to each SED plot, the panels in the right column show posteriors (probability density) for $\mstar$, dust attenuation, and mass-weighted (MW) age for each galaxy (w/wo MIRI). The dashed lines indicate 16\%, 50\%, and 84\% intervals.
}}
	\label{fig:SED_comp}
\end{figure*}

\subsection{Stellar mass estimation}
\label{subsec:mstar}
We performed two suites of SED fitting with BAGPIPES \blue{~\citep{Carnall:2018}} w/wo MIRI photometry, and derived $\mstar$ for all galaxies, dubbed as $M_{\star, \rm MIRI}$ and $M_{\star,\rm noMIRI}$, respectively.
We used the 2016 update of the stellar population synthesis model by \cite{Bruzual:2003} (age $\in \rm[0.03,10]~Gyr$ and metallicity $Z/Z_\odot\in[0.01,2.5]$), 
a delayed exponentially declining star formation history (timescale $\rm \tau \in[0.01,10]~Gyr$), dust models in \cite{Calzetti:2000} for young and old 
stellar population separately (divided by 0.01~Gyr; $A_V \in[0,5]$), and nebular emission constructed following \cite{Byler:2017} with 
ionization parameter $\log U\in[-5,-2]$ to run BAGPIPES.

We also used \textsc{fast++}\footnote{\url{https://github.com/cschreib/fastpp}} to derive $\mstar$ for massive galaxies, adopting the stellar population synthesis model in
\cite{Bruzual:2003} with stellar ages $\in \rm[0.1,10]~Gyr$, metallicity $Z/Z_\odot\,=\,[0.2, 0.4, 1.0, 2.5]$, 
a delayed star formation history (timescale $\rm \tau \in[0.1,100]~Gyr$),
the dust extinction model of \cite{Calzetti:2000}. Stellar masses from both methods agree within $\sim$0.3 dex.

Mass completeness estimates using maximally old galaxy templates are shown in Appendix~\ref{appendix:mass_complete}. The F444W depth ensures completeness for $\mstar > 10^{9}M_{\odot}$ galaxies up to $z \sim 8.5$. All $\mstar > 10^{10}M_{\odot}$ galaxies in our sample are detected in MIRI/F770W (S/N $>$ 4). However, completeness is not only color (age) dependent but also morphology (and dust extinction) dependent. We acknowledge that we might run into completeness issues at the lowest masses and highest redshifts.

Our sample includes 23001 galaxies at $2.5 < z < 8.5$ with S/N$_{\rm F444W} > 7$. Of these, 11450 galaxies have F770W detections with S/N $>$ 2, including all 881 massive galaxies ($\mstar > 10^{10} M_{\odot}$).

\subsection{Selection of ``Little Red Dots''}
Many photometrically selected massive galaxies in previous JWST/NIRCam studies are candidate Type-I AGNs with broad emission lines and point-like morphologies (``Little Red Dots'', LRDs). Recent studies based on JWST/MIRI suggest most LRDs have low masses with $\mstar \sim 10^{9-9.5} M_\sun$, although some may be quite massive. 
Using our large sample with MIRI photometry, we systematically select LRDs and examine their impact on massive galaxy selection and number density in the early universe. 
We adopt the same LRD criteria from  \citet{Labbe2025}, which require both red colors and compactness. We performed 2D light-profile fits in F444W to identify point-source-dominated galaxies (L. Chen et al. in prep). Following \citet{Weibel2024}, we required $S/N > 3$ in all color-selection bands for a clean LRD sample.

Of 23001 galaxies, 93 are classified as LRDs. This is consistent with \citet{Weibel2024}, who found 105 LRDs within the same two PRIMER fields. Due to uncertainties in $\mstar$ estimates and the elusive nature of LRDs, we exclude these 93 LRDs from our primary sample. 

Stellar masses for LRDs are derived identically to normal galaxies, i.e., based on solely stellar population templates. These $\mstar$ values can be considered upper limits~\citep[see also, e.g.,][]{Baggen:2024LRD}. To study MIRI's effect on LRD $\mstar$ estimates, we also ran the same two SED-fitting suites, with and without including MIRI.

\begin{figure*}
	\centering
	\includegraphics[scale=0.40]{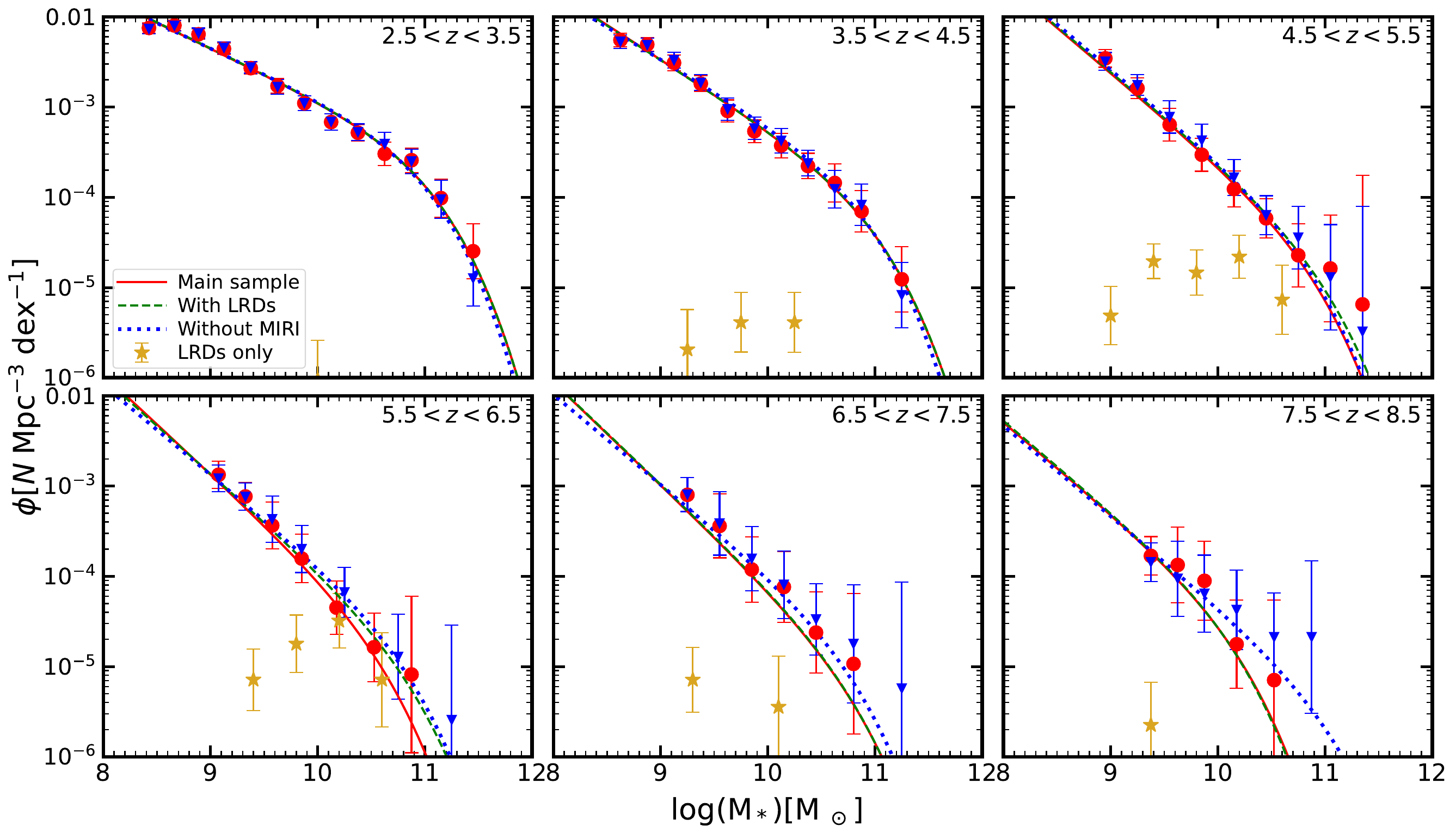} 
	\caption{\small{\textbf{Stellar mass functions derived with (red circles) and without MIRI (blue triangles) at $z \sim 3-8$.} The filled red circles and blue triangles denote the data points with $\mstar$ derived with and without MIRI data, respectively. The red and blue lines show best-fit Schechter functions (before convolution) for the data points with and without MIRI data. The green dashed lines show best-fit Schechter functions including our primary sample and LRDs. The orange pentagrams show the stellar mass distribution of LRDs only, for which $\mstar$ are estimated using solely stellar templates. 
}}
	\label{fig:SMF}
\end{figure*}

\section{Results}\label{sec:results}
\subsection{The importance of MIRI in constraining stellar masses at $z > 5$}
Figure~\ref{fig:color_color} shows the distribution of massive galaxies with $M_{\star, \rm MIRI} > 10^{10.3} M_{\odot}$ or $M_{\star, \rm noMIRI} > 10^{10.3} M_{\odot}$ at $z > 5$ in the F277W $-$ F444W vs. F150W $-$ F277W diagram. 
The number of $M_{\star} > 2 \times 10^{10} M_{\odot}$ galaxies drops significantly when MIRI is included, 31 vs. 68 without MIRI. Most MIRI-selected massive galaxies fall outside the green shaded region. This region was originally used to select massive galaxies with V-shaped SEDs, which are now known mostly actually LRDs or brown dwarfs \blue{when combined with an additional compactness criteria}~\citep{Labbe:2023,Labbe2025,Greene:2024,Barro:2024,Kocevski:2025}.

Most MIRI-selected massive galaxies  cannot be identified as LRDs, with an LRD fraction of 13\% (4/31) vs. 32\% (22/68) without MIRI.
Without MIRI, many objects with red F277W $-$ F444W and blue F150W $-$ F277W are misidentified as massive $z > 5$ galaxies; most of them are disqualified when MIRI is added due to overestimation of $\mstar$ and/or being LRDs. 


Figure~\ref{fig:mass_diff} compares $\mstar$ estimates with/without MIRI at fixed redshifts. For $M_{\star,\rm noMIRI} > 10^{10} \msun$ galaxies at $z > 5$, $\mstar$ is systematically overestimated by $\sim 0.4$ dex when MIRI is excluded-where the longest NIRCam filter (F444W) probes only rest-frame $< 1$ $\mu$m. Examination of the SED fits reveals two primary reasons for the overestimation (Figure~\ref{fig:SED_comp}). One is the age-attenuation degeneracy: Without MIRI, best-fit SEDs favor younger ages and higher dust attenuation, leading to an overestimation of $\mstar$. The other is emission-line contamination: When strong lines fall in long-wavelength filters(e.g., F444W or F356W), the best-fit SEDs tend to overestimate the stellar continuum and $\mstar$. 
In many cases, the overestimation of $\mstar$ involves both factors (also see, e.g., \citealt{Papovich:2023}). 
The effect of MIRI on the $\mstar$ estimate of lower-mass galaxies is less pronounced, likely due to intrinsically younger ages and bluer SEDs reducing the age-attenuation degeneracy.

While no systematic $\mstar$ differences are found at $z < 5$, some galaxies at these redshifts can still be overestimated without MIRI. Figure~\ref{fig:SED_comp} shows SED-fitting results for two example galaxies at $z \sim 3$, which are characterized by much younger and dustier SEDs with lower $\mstar$ when MIRI is included.

\begin{figure*}
	\centering
     \includegraphics[scale=0.4]{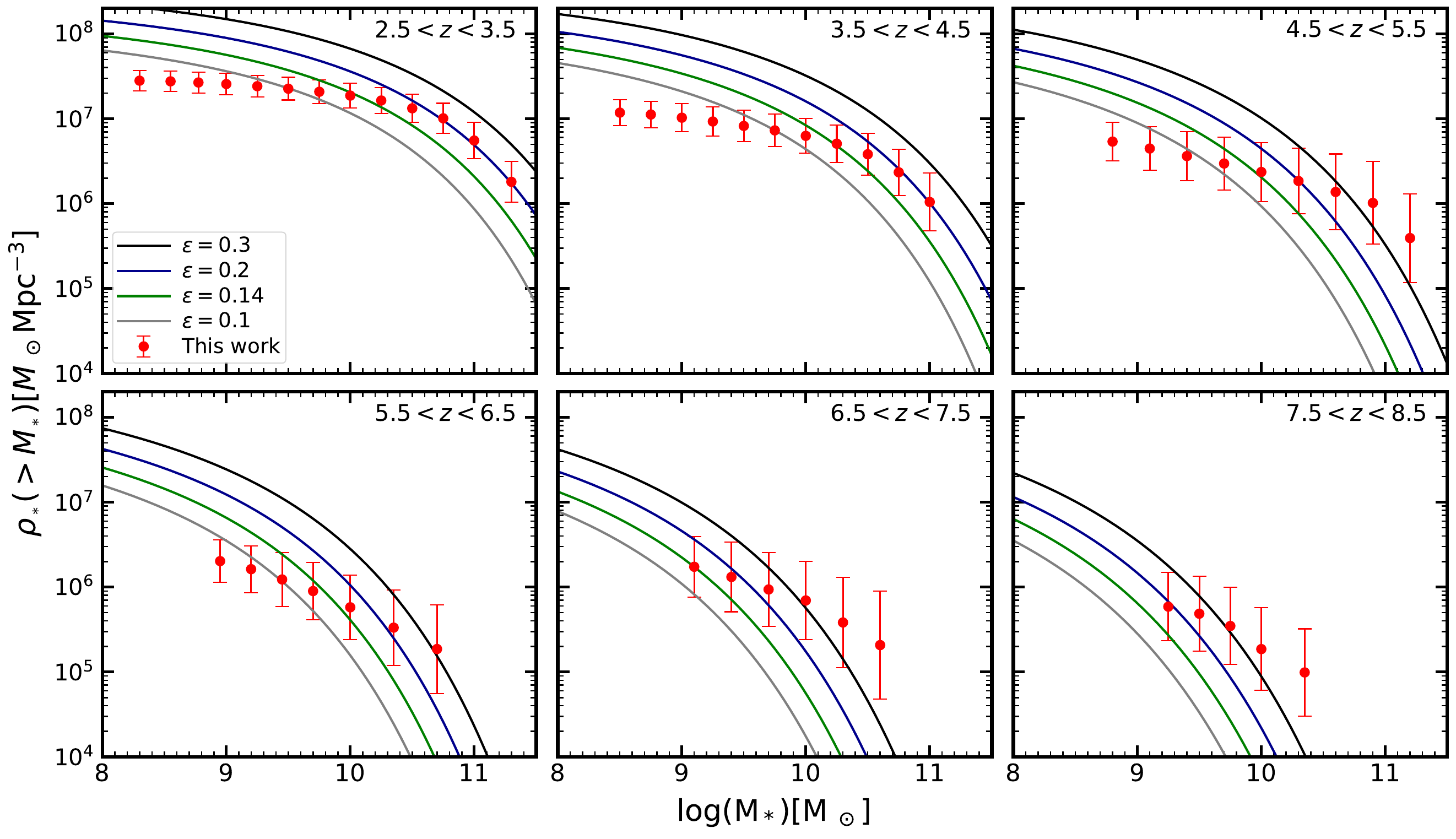}
	\caption{\small{\textbf{Observed cumulative stellar mass densities compared to $\Lambda$CDM predictions for varying $\epsilon$.}
 The filled red circles denote the observed cumulative stellar mass densities based on our primary sample. The error bars indicate 1$\sigma$ uncertainty that includes both Poisson errors and cosmic variance. The colored lines show predictions for four $\epsilon$ values given the assumed halo mass functions. 
 }}
	\label{fig:CSMF}
\end{figure*}

\begin{table}
\caption{Best-fit Schechter function parameters for SMFs with MIRI photometry \label{Tab:parameters}}
\begin{tabular}{cccc}
\hline\hline
Redshift  & $\log ({M_{\star}})$  & ${\Phi ^ * }\times {10^5}$ & ${\alpha }$\\
 & $({M_ \odot })$& ${\rm{Mp}}{{\rm{c}}^{ - 3}}{\rm{de}}{{\rm{x}}^{ - 1}}$ & \\
\hline
3 & [11.0] & $13.6 \pm 1.5$ & $-1.57 \pm 0.06$\\
4 & $10.94 \pm 0.17$ & $4.49 \pm 0.26$ & $-1.77 \pm 0.06$ \\
5 & $10.81 \pm 0.22$ & $1.44 \pm 0.25$ & [-2.0] \\
6 & $10.54 \pm 0.25$ & $1.06 \pm 0.31$ & [-2.1]\\
7 & $10.74 \pm 0.41$ & $0.41 \pm 0.11$ & $-2.13 \pm 0.20$ \\
8 & $10.12 \pm 0.53$ & $1.51 \pm 0.52$ & [-2.0] \\
\hline
\end{tabular}
\end{table}

\subsection{The stellar mass function of galaxies at $z = 3-8$}
 
To quantify the effect of MIRI photometry on selecting high$-z$ massive galaxies, we derive SMFs using $\mstar$ with/without MIRI.
To account for Eddington bias and recover intrinsic SMFs, we convolve the Schechter function $\Phi(M)\rm{d(}\log M)  = \ln (10) \times {\Phi ^ * } \times \exp \left[ - {10^{\log (M/M_{\star}) }}\right] \times {\left[{10^{\log (M/M_{\star}) }}\right]^{\alpha  + 1}}{\rm{d(}}\log M)$ with a Gaussian kernel ($\sigma  = 0.2$) including a Lorenztian ``wing'' to suppress outliers on the massive end~\citep{Davidzon:2017,Adams:2021,Weaver:2023B}. 
We then perform ${\chi ^2}( = \sum {({\rm{resi}}{{\rm{d}}^2}/{\sigma ^2})} )$ minimization using Scipy v1.8.1 \citep{Gommers2022}, with errors ${\sigma ^2} = \sigma _{{\rm{cv}}}^2 + \sigma _{\rm{N}}^2$ including the cosmic variance ${\sigma _{{\rm{cv}}}}$ given by the “cosmic variance cookbook” \cite{Moster:2011} and the Poisson noise ${\sigma _{{\rm{N}}}}$ derived from the intrinsic number counts (estimated from the best-fit Schechter function before convolution). To avoid complicating the SMF measurement, we do not include the previously mentioned systematic biases from SED fitting to the error bars. The priors used are $[ - 1.5 - 0.1 \times z, - 1.2 - 0.1 \times z]$ for $\alpha$, and $[9.0,11.0]$ for ${\log {M_{\star}}}$. Best-fit SMF parameters with MIRI photometry are listed in Table~\ref{Tab:parameters}.

Figure~\ref{fig:SMF} compares SMFs derived from our primary sample~(i.e., excluding LRDs) w/wo MIRI. SMFs with MIRI match those without at $z < 5$, but are lower at the massive end for $z > 5$. This is consistent with individual galaxy differences shown in Figure~\ref{fig:mass_diff}.

In order to have some rough estimation on the effect of including LRDs in the SMF estimation, we also show the stellar mass distribution of LRDs in Figure~\ref{fig:mass_diff}. We emphasize here that the $\mstar$ of LRDs are derived in the same way as the other normal galaxies, which are based solely on stellar templates with MIRI photometry. The contribution of LRDs to the SMF is most significant at $z \sim 4.5-6.5$, decreasing at higher/lower redshifts. Their characteristic $\mstar \sim 10^{9.5-10.5} \msun$ aligns with studies using more complicated galaxy/AGN templates~\citep{Baggen:2024LRD,PerezGonzalez:2024,Akins:2024}.

\subsection{An increasing baryon conversion efficiency towards higher redshifts}
Using photo$-z$s and $\mstar$ derived with MIRI, we calculate the cumulative stellar mass density for $M_{\star} > 10^{9}M_{\odot}$ galaxies at $2.5 < z <9$ (Figure~\ref{fig:CSMF}).  
Figure~\ref{fig:CSMF} also shows expected cumulative stellar mass densities assuming a constant integrated efficiency of converting baryons to stars ($\epsilon$), 
$\rho_{\star}(>\mstar)=\epsilon \rho_{\rm m}(>\mstar/\fbary)$, where $\fbary = \Omega_{\rm b}/\Omega_{\rm m}$ is the cosmic baryon fraction, and $\rho_{\star}$ and $\rho_{\rm m}$ are cumulative stellar and DM halo mass densities, respectively. The cumulative DM halo mass density more massive than $M_{\rm h} = \mstar/(\fbary\epsilon)$ is derived from the DM halo mass function in~\cite{Watson:2013}. Using the DM halo mass functions from the Bolshoi-Planck and 
MultiDark-Planck \lcdm~ cosmological simulations~\citep{RodrguezPuebla:2016}, as used in~\cite{Chworowsky2024}, yields similar results.

Two prominent features emerge from the comparison of the cumulative stellar mass densities between observations and model predictions. 
Firstly, the relatively low abundance of massive galaxies at $z \sim 4-8$ obtained by including MIRI is fully consistent with \lcdm\ predictions without requiring extreme $\epsilon$. Secondly, $\epsilon$ 
increases towards higher stellar masses and higher redshifts. However, even for the most massive galaxies at $z \sim 8$, 
only a moderately enhanced efficiency of $\epsilon \sim 0.3$ is required, in comparison to $\epsilon \sim 0.1-0.2$ for typical galaxies at $z \lesssim 3$~\citep{Bregman:2007,Behroozi:2019,ZhangZ:2022}.

\begin{figure*}
	\centering
	\includegraphics[scale=0.5]{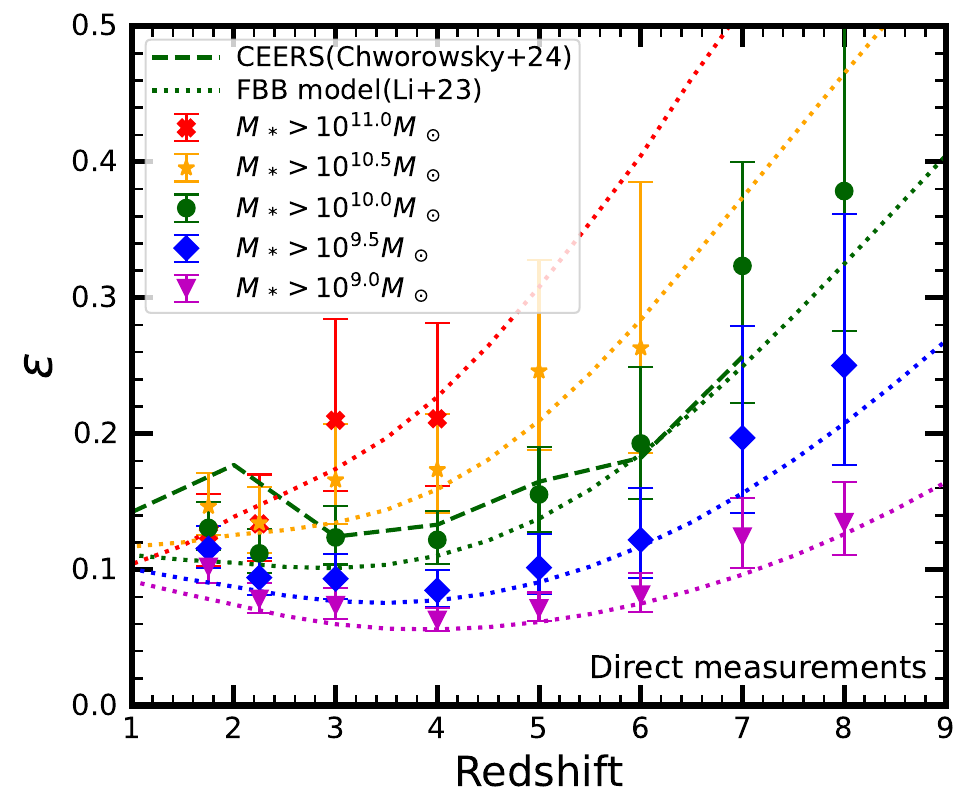}
    \includegraphics[scale=0.5]{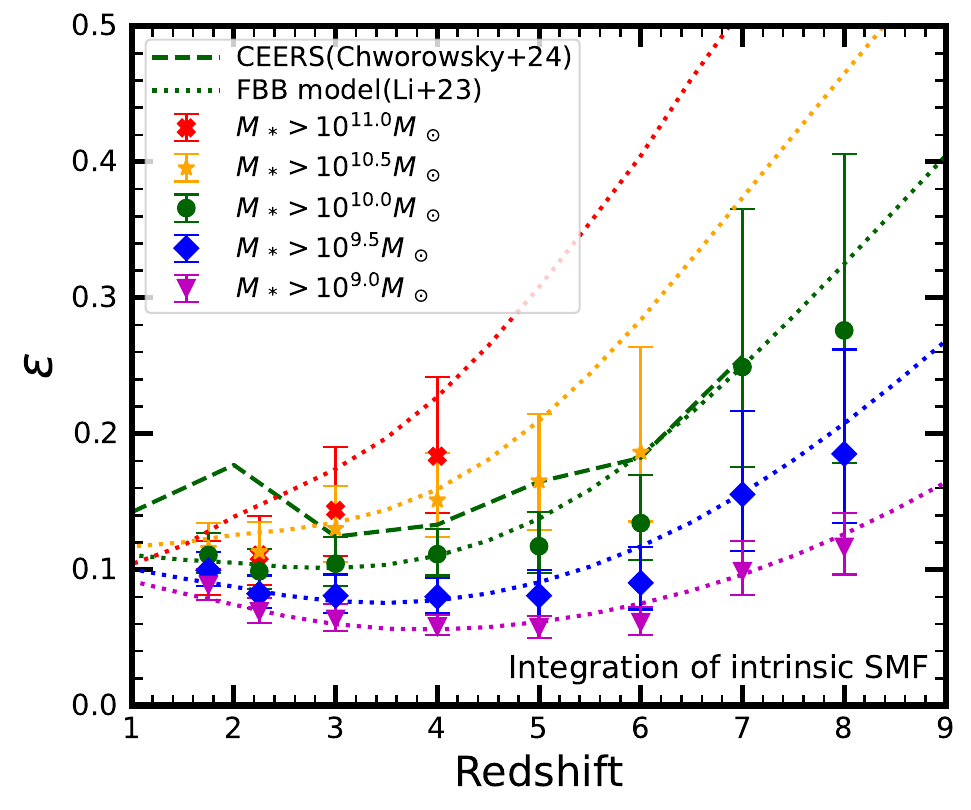}
\caption{\small{\textbf{Redshift evolution of the observed $\epsilon$ and comparison to models.} 
The data points show best-matched $\epsilon$ at different mass cuts and redshifts, based on $\mstar$ estimate for the primary sample (excluding LRDs) with MIRI photometry. The difference is that $\epsilon$ shown in the left panel is derived from observed cumulative number densities of galaxies, while in the right panel they were obtained by integrating the intrinsic SMF (before convolution with a Gaussian kernel to account for the Eddington bias). The dotted lines show FFB models with $\epsilon_{\rm max} = 0.2$ that best matches our data~\citep{LiZ:2023}. The dashed line shows CEERS results for $\mstar > 10^{10} \msun$ galaxies based on NIRCam data only~\citep{Chworowsky2024}.
}}
	\label{fig:e_evol}
\end{figure*}


Figure~\ref{fig:e_evol} shows $\epsilon$ evolution with mass and redshift, derived from both observed stellar mass densities (left panel) and intrinsic SMF (right; correcting Eddington bias; ~\citep{ChenY:2023}). 
At all masses, $\epsilon$ remains quite flat at low redshifts, which is consistent with previous studies on a rather constant $\epsilon$, or equivalently a constant stellar-to-halo mass ratio at $0 < z \lesssim 3$~\citep{Behroozi:2019}. At $z \gtrsim 3$, $\epsilon$ increases rapidly with redshift. This transition occurs earlier for less massive galaxies, with $z \sim 3$ for $\mstar > 10^{11} M_{\odot}$ and $z \sim 5$ for $\mstar > 10^{10} M_{\odot}$.

\section{Discussion}
\label{sec:discussion}
The discoveries of this paper have important implications for the studies of high$-z$ galaxies. Here we briefly discuss a few aspects 
regarding massive galaxy selection, the role of LRDs, stellar mass estimation, and  baryon-to-star conversion efficiency at $z \sim 3-8$.

Photometric selection of high-redshift massive (old or dusty) galaxies relies on Balmer/4000 \AA~break or submillimeter emission. 
Based on the presence of the Balmer break and blue UV continuum of massive (star-forming) galaxies, \cite{Labbe:2023} proposed selecting $z > 6$ massive galaxies using red F277W-F444W and blue F150W-F277W colors. 
However, Figure~\ref{fig:color_color} shows that MIRI-confirmed massive galaxies have redder UV slopes than this criterion, which are expected for massive (and dusty) star-forming galaxy templates.

We robustly demonstrate MIRI's critical role in constraining  $\mstar$ for $z > 5$ massive galaxies. It calls for great caution against the massive end of the SMF derived from the photometry with coverage only up to IRAC 4.5$\mu$m or to NIRCam/F444W, which tend to be overestimated. These features are distinctively illustrated in Figure~\ref{fig:SMF} via the comparison between the SMFs derived with and without MIRI (also see Figure~\ref{efig:SMF} for comparison with other studies), which are in good agreement  at $\mstar < 10^{10}\msun$ across all redshifts but differ at $\mstar > 10^{10}\msun$ and $z > 5$.

One common concern in current SMF estimation at high redshifts is the $\mstar$ of LRDs, or strong emission-lines galaxies in general. MIRI largely suppress the overestimation of $\mstar$ of strong line-emitters at $z \sim 3-8$, for which strong optical emission lines fall mainly in the NIRCam bands, as can be seen in the example SEDs in Figure~\ref{fig:SED_comp}. Assuming a purely stellar origin, LRDs have (maximum) stellar masses of $\mstar \sim 10^{9-10.5} \msun$. These values are  consistent with recent studies on the stellar masses of LRDs using more complicated galaxy and AGN templates~\citep[see, e.g.,][]{PerezGonzalez:2024}. It is also consistent with recent clustering analysis, suggesting that LRDs are not among the most massive galaxies at their epochs~\citep{Pizzati:2024}.


\begin{figure*}
	\centering
	\includegraphics[scale=0.4]{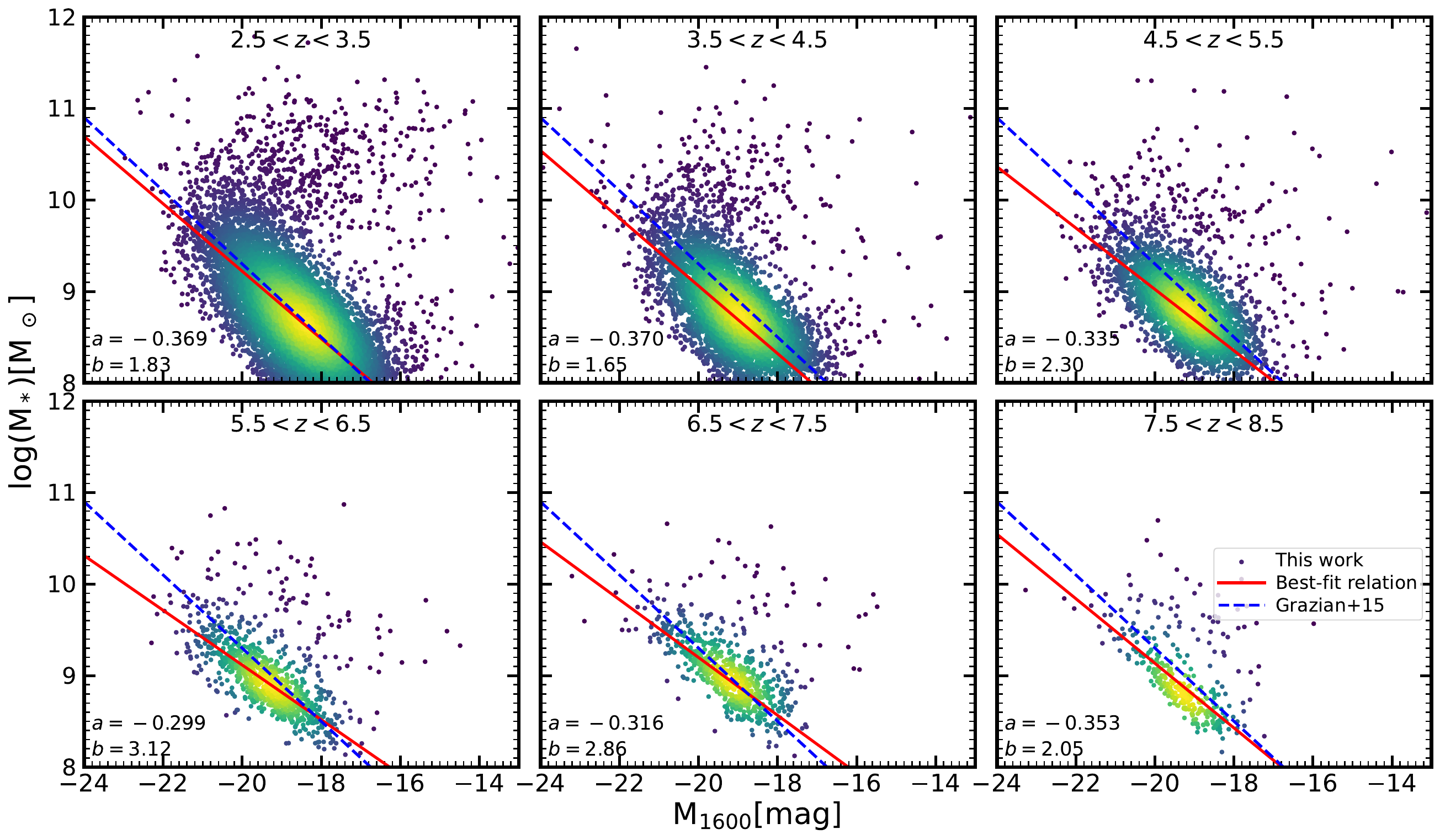}
	\caption{\small{\textbf{$\mstar$ vs. absolute UV magnitude ($M_{\rm 1600}$) for our primary galaxy sample.} 
The red solid lines show best-fit linear relations per redshift bin, while previous estimates from \cite{Grazian:2015} are shown by the blue dashed lines. The best-fit parameters for the linear $\mstar-M_{\rm 1600}$ relations (log$\mstar = a \times M_{\rm 1600} + b$) are indicated at the bottom-left corner of each panel. 
}}
	\label{fig:M1600_Mstar}
\end{figure*}

Our finding of moderately increasing $\epsilon$ with redshift at $z = 3-8$ agrees with recent studies using JWST/NIRCam data after accounting for the contamination of LRDs~\citep{WangY:2023,Chworowsky2024}. 
Using our mass-complete samples, we further show $\epsilon$ also increases with $\mstar$ (up to $\mstar \sim 10^{11} \msun$ at $3 < z < 8$), a trend that has been known for galaxies at lower redshifts ($0 < z < 3$)~\citep[see, e.g.,][]{Behroozi:2019}.


Debate persists on whether $\epsilon$ increases with redshift. Some studies using UV luminosity functions argue against higher $\epsilon$ at high$-z$~\citep[e.g.,][]{Donnan:2025}. While a thorough investigation of the UV luminosity function is beyond the scope of this paper, we caution that converting $M_{\rm UV}$ to $\mstar$ via a linear relationship underestimates massive galaxy number densities. Pre-JWST studies showed that most $\mstar > 10^{10} \msun$ galaxies at $z \gtrsim 3-4$ are fainter in the UV than predicted by linear $\mstar-\muv$ relation~\citep{WangT:2016a} calibrated for low-mass, UV-bright galaxies. We extend this to higher redshifts using PRIMER galaxies (Figure~\ref{fig:M1600_Mstar}). Massive galaxies at $z \sim 3-8$ deviate significantly from linear $\mstar-M_{\rm 1600}$ relations. These massive galaxies would have underestimated $\mstar$ or be missed entirely in Lyman-break galaxy (LBG) selections due to low $M_{\rm UV}$ and/or high extinction. The same reasoning can also explain why earlier studies based on UV-selected galaxy samples suggested decreasing $\epsilon$ at $z\sim 3-8$~\citep{Behroozi:2019}, which likely missed a significant fraction of massive galaxies.

On the theoretical side, the increasing trend of $\epsilon$ with redshift and stellar masses is consistent with the feedback-free starburst model (FFB model,~\citealt{Dekel:2023,LiZ:2023}). The FFB model posits that high gas density, intermediate metallicity, short free-fall time, and rapid cooling in massive high$-z$ galaxies suppress stellar feedback, leading to higher $\epsilon$.
Furthermore, FFB is expected to be associated preferentially with halos above a threshold mass that declines with redshift. 
As shown in Figure~\ref{fig:e_evol}, the trends of an increasing $\epsilon$ with redshifts and stellar masses are generally consistent with the FFB model assuming $\epsilon_{\rm max} = 0.2$.  While these results are based on our primary sample after excluding LRDs, including LRDs does not significantly change our results. Adding LRDs will slightly increases $\epsilon$ at $z \sim 5-6$, but the trend of increasing $\epsilon$ with redshift and $\mstar$ remains and matches FFB predictions. 
Although the current study is limited to $z \lesssim 9$, the trend shown in Figure~\ref{fig:e_evol} indicates an even larger $\epsilon$ for galaxies at $z > 9$, consistent with the 
high abundance of UV-luminous galaxies at $z > 9$ and predictions of FFB models. 

Current estimates of $\epsilon$ are largely based on the abundance matching between galaxies and dark matter (DM) halos. More direct constraints on DM halo masses, such as through clustering analysis, are needed to further consolidate these results.

\section{conclusion}
\label{sec:conclusion}
This study leverages deep multiwavelength data from the JWST/PRIMER survey, combined with NIRCam and MIRI photometry, to investigate the abundance and properties of massive galaxies at $3 < z < 8$. Our key findings include:
\begin{itemize}
\item Stellar mass estimates for massive galaxies ($M_{\star, {\rm noMIRI}} > 10^{10}\msun$) at $z > 5$ are systematically overestimated by $\sim 0.4$ dex when relying solely on NIRCam data. MIRI photometry breaks degeneracies in SED fitting (e.g., age-dust attenuation, emission-line contamination), reducing spurious identifications of extremely massive galaxies. 
\item The inclusion of MIRI lowers the massive-end SMF at $z > 5$ compared to previous estimates without MIRI, largely resolving tensions with $\Lambda$CDM predictions. Previous overestimates stemmed from limited rest-frame near-infrared coverage and contamination from strong line-emitters (including LRDs). The SMFs derived with MIRI align with theoretical expectations, requiring only moderate baryon-to-star conversion efficiencies for massive galaxies, $\epsilon \sim 0.3$ for galaxies with $\mstar > 10^{10} \msun$ at $z = 8$. 
\item We put constraints on the abundances and masses of LRDs. Their stellar masses are in the range of $\mstar \sim 10^{9-10.5} \msun$ assuming a purely stellar origin, confirming that they do not dominate the high-mass end of the SMF. Their contribution to the massive galaxy population is most significant at $4.5 < z < 6.5$, which rapidly decreases towards both high and lower redshifts.
\item 
We reveal that the baryon-to-star conversion efficiency ($\epsilon = \mstar/M_{\rm h}/f_{\rm b}$) remains rather constant at $z < 3$, above which enters a phase of rapid growth with redshift. This transition takes place at higher redshifts for less massive galaxies. This trend matches predictions from feedback-free starburst (FFB) models, where high gas densities and the associated short free-fall time in early massive halos enables rapid, efficient star formation.
\item
We show that massive galaxies at $z \sim 3-8$ deviate significantly from linear  $\mstar-\muv$ relations calibrated for low-mass, UV-bright populations, highlighting biases in UV-selected samples and the approach of converting UV luminosities to $\mstar$ based on a linear $\mstar-\muv$ relation.
\end{itemize}
The study underscores the necessity of MIRI data for robust high$-z$ galaxy studies, particularly for  $\mstar > 10^{10} \msun$ galaxies. The rising baryon conversion efficiency suggests early massive galaxies formed stars under unique physical conditions distinct from their low$-z$ counterparts. 

One limitation of this work is that it is still based on mainly photometric samples, although the accuracy of photo-z estimate for high$-z$ galaxies is significantly improved compared to pre-JWST studies. Complete spectroscopic samples of massive galaxies at $z\sim 4-8$, including more reliable stellar mass estimates of LRDs, are key to consolidating these results and providing a clear answer regarding the number density of massive galaxies and the baryon-to-star conversion efficiency in the early Universe.




\section{Acknowledgments}
We thank the reviewer for carefully reading our manuscript and providing very constructive comments, which helped to improve significantly the quality of the paper.
 T.W. acknowledges support by National Natural Science Foundation of China (Project No. 12173017 and Key Project No. 12141301), Scientific Research Innovation Capability Support Project for Young Faculty (Project No. ZYGXQNJSKYCXNLZCXM-P3), National Key R\&D Program of China (Project No. 2023YFA1605600), and the China Manned Space Project with grant No. CMS-CSST-2025-A04. L.Z. acknowledges the support from the National Natural Science Foundation of China (Project No. 13001103). Y.J.W. acknowledges support by National Natural Science Foundation of China (Project No. 12403019) and Jiangsu Natural Science Foundation (Project No. BK20241188). We are grateful for Dr. Zhoujian Zhang for his help with the synthetic {\it JWST} photometry of brown dwarfs. This work is based [in part] on observations made with the NASA/ESA/CSA James Webb Space Telescope. The data were partly obtained from the Mikulski Archive for Space Telescopes (MAST) at the Space Telescope Science Institute, which is operated by the Association of Universities for Research in Astronomy, Inc., under NASA contract NAS 5–03127 for JWST. These observations are associated with programme 1837, 1727, 3990, and 1840.
The specific data from MAST analyzed can be accessed via \dataset[doi:10.17909/xcqt-gw25]{https://doi.org/10.17909/xcqt-gw25}.
 
 We are grateful to Dr. Gabriel Brammer for sharing the {\it JWST}/NIRCam and {\it HST} data. Part of the {\it JWST}/NIRCam and {\it HST} data products presented here were retrieved from the Dawn JWST Archive (DJA). DJA is an initiative of the Cosmic Dawn Center (DAWN), which is funded by the Danish National Research Foundation under grant DNRF140.

\facilities{HST, JWST, Subaru, Vista, Spitzer, CFHT}

\appendix
\restartappendixnumbering
\section{JWST data reduction}
\label{appendix:MIRI}

\begin{figure*}
	\centering
    \includegraphics[width=1\textwidth]{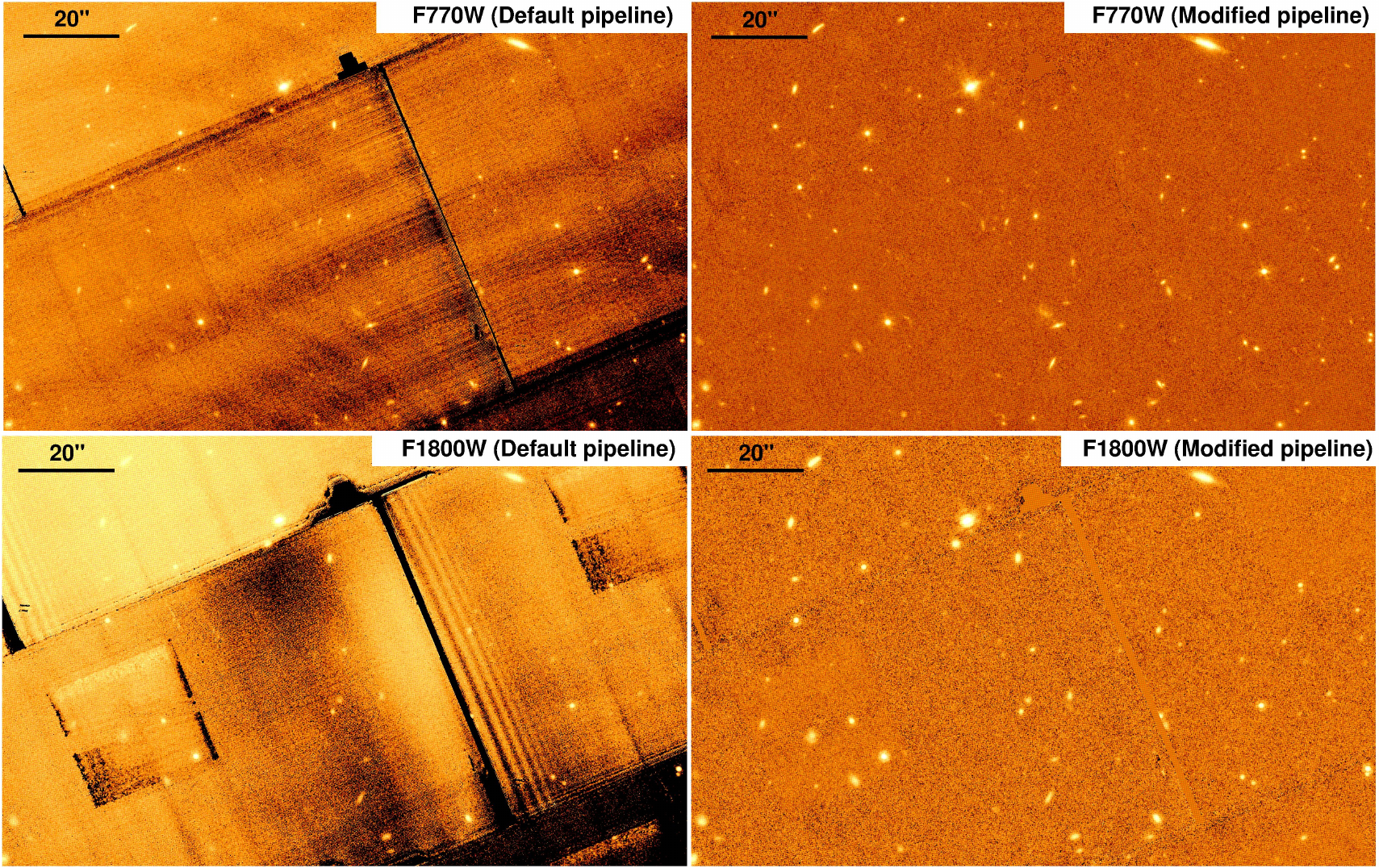} 
	\caption{\small{\textbf{Comparison of MIRI images reduced with default and our modified pipelines in PRIMER-COSMOS.} Left: MIRI F770W and F1800W images reduced with default JWST calibration pipeline (level 3 science images from the MAST archive). Right: MIRI F770W and F1800W images reduced with modified pipelines.
}}
	\label{efig:MIRI-images}
\end{figure*}

The JWST/NIRCam and MIRI images were reduced following the JWST Calibration Pipeline v1.13.4 \citep{Bushouse2024}. 
Since the default output of the default Pipeline still has a few defects (see, e.g., \cite{Bagley2023,YangG:2023}), we use our custom-made steps to improve the quality of data reduction, which is summarized in Table~\ref{tabA1}. More detailed information about our pipeline will be shown in a forthcoming paper (H. Sun et al., in preparation). Here we only list the major revisions as follows.
Firstly, in stage 2 of the reduction of MIRI data, we replace the default flat fields  with new ones constructed per pointing by stacking rate maps with source-masking from the adjacent 5 pointings that were produced from stage 1. 
Secondly, after stage 2, in addition to subtract remaining ``stripe-like'' noise patterns following \citet{Bagley2023} for NIRCam and \citet{YangG:2023} for MIRI, we use photutils v1.5.0 \citep{Bradley:2022} to subtract the background from each "cal" image (the output from stage 2 of the pipeline). This is expected to solve the problem of uneven background, and to improve the performance of the default source detection algorithm in stage 3. 
Thirdly, when using the `TweakReg' routine in stage 3 to perform astrometry calibration, we replace the default Gaia catalog, which has too few sources, 
with a reference catalog based on HST/WFC3 F160W or HST/ACS F814W images from Grizli Image Release v7.0. 
Figure~\ref{efig:MIRI-images} compares example MIRI images reduced with the default and our modified pipeline.
We derive the depth of these images for point sources by measuring the fluxes in empty apertures and calculate the standard deviation of them. 
Using the aperture with the highest S/N for each band, we find that the 3$\sigma$ depth for the F444W, F770W and F1800W images are 29.2, 26.7 and 23.8 mag, respectively.

\begin{table*}
\centering
\begin{minipage}[center]{\textwidth}
\centering
\caption{MIRI data reduction procedures.\label{Tab:MIRI_pipeline}}
\begin{tabular}{ll}
\hline\hline
Procedure  & Description\\
\hline
Stage 1$^a$ & Detector-level corrections\\
Source masking & Mask source-affected pixels before subtraction of flat-field and stripes.\\
Flat-field construction$^b$ & Construct super-sky flat per pointing\\
Stage 2 & Instrumental corrections and calibration\\
Stripe subtraction & Subtract residual stripes following \citep{Bagley2023,YangG:2023} \\
Background subtraction & Subtract background per ``cal'' image\\
TweakReg & Compute WCS corrections \\
SkyMatch$^a$ & Calculate sky values in overlaps \\
Outlier Detection$^a$ & Reject the pixels affected by cosmic rays or other artifacts\\
Resample$^a$ & Resample and combine images  \\
\hline
\end{tabular}
\begin{flushleft}
{\sc Note.} --- 
($a$) These procedures are performed with default parameters. 
($b$) This step is only used for JWST/MIRI images.
\end{flushleft}
\end{minipage}
\label{tabA1}
\end{table*}

\section{Comparison with Reference Photometry}
\label{appendix:phot_comp}
\begin{figure*}
	\centering
	\includegraphics[scale=0.5]{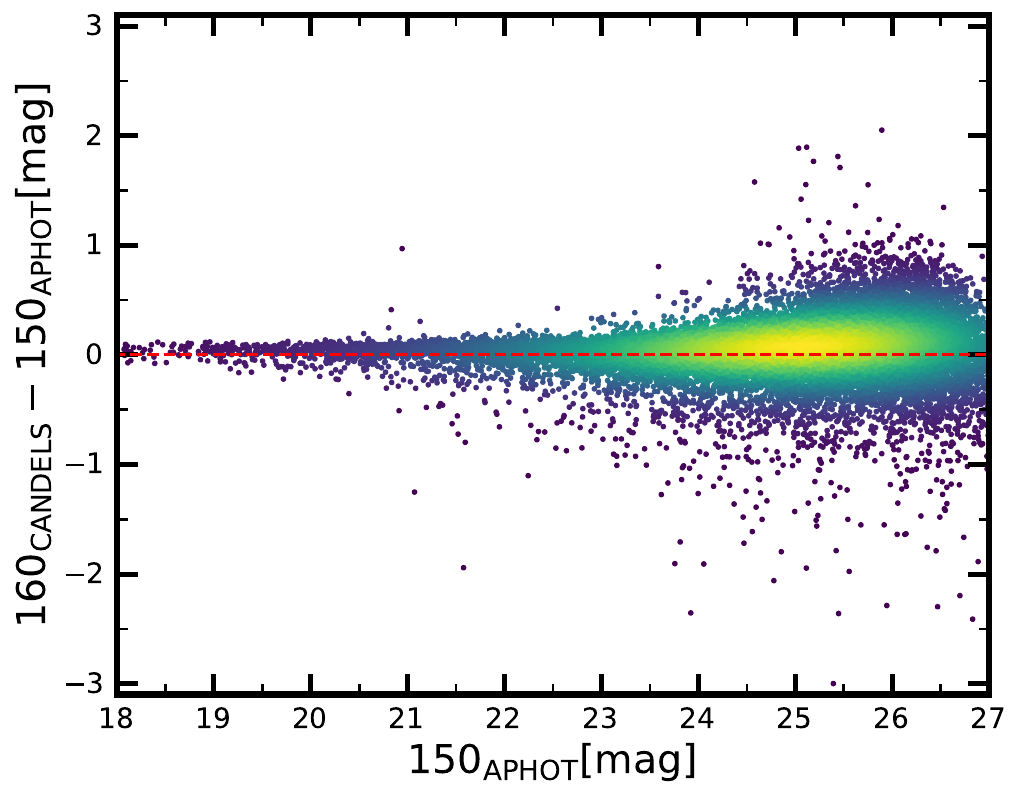}
    \includegraphics[scale=0.5]{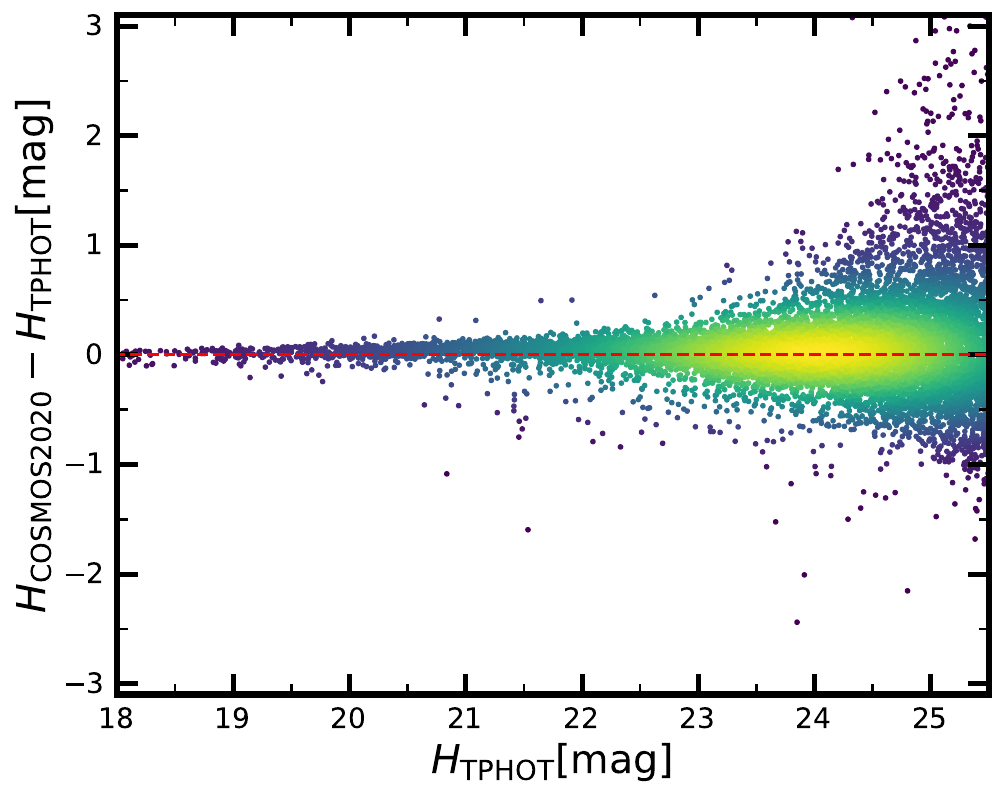}
\caption{\small{\textbf{Comparison of our JWST/NIRCam F150W (aperture photometry using APHOT) and VISTA/H photometry (de-blended photometry with TPHOT) with previous measurements by other teams.} 
The color maps indicate the local density of data points. The {\it HST}/FI60W photometry is obtained from the CANDELS survey \citep{Nayyeri2017}, while the {\it VISTA}/{\it H} band photometry is obtained from the COSMOS2020 catalog \citep{Weaver2022}. The red dashed lines denote the median values of the magnitude differences, which is close to zero across a wide magnitude range.
}}
	\label{efig:phot_compare}
\end{figure*}

Figure~\ref{efig:phot_compare} compares our {\it JWST}/NIRCam F150W (APHOT) and VISTA/{\it H} (TPHOT) photometry with CANDELS ({\it HST}/F160W,~\citealt{Nayyeri2017}) and COSMOS2020 (VISTA/{\it H};~\citealt{Weaver2022}). Both APHOT (high-resolution) and TPHOT (low-resolution) fluxes agree well with previous results, with no systematic offsets.

\section{The impact of MIRI photometry on photo-z estimate}
\label{appendix:photz_comp}
\begin{figure*}
	\centering
	\includegraphics[scale=0.7]{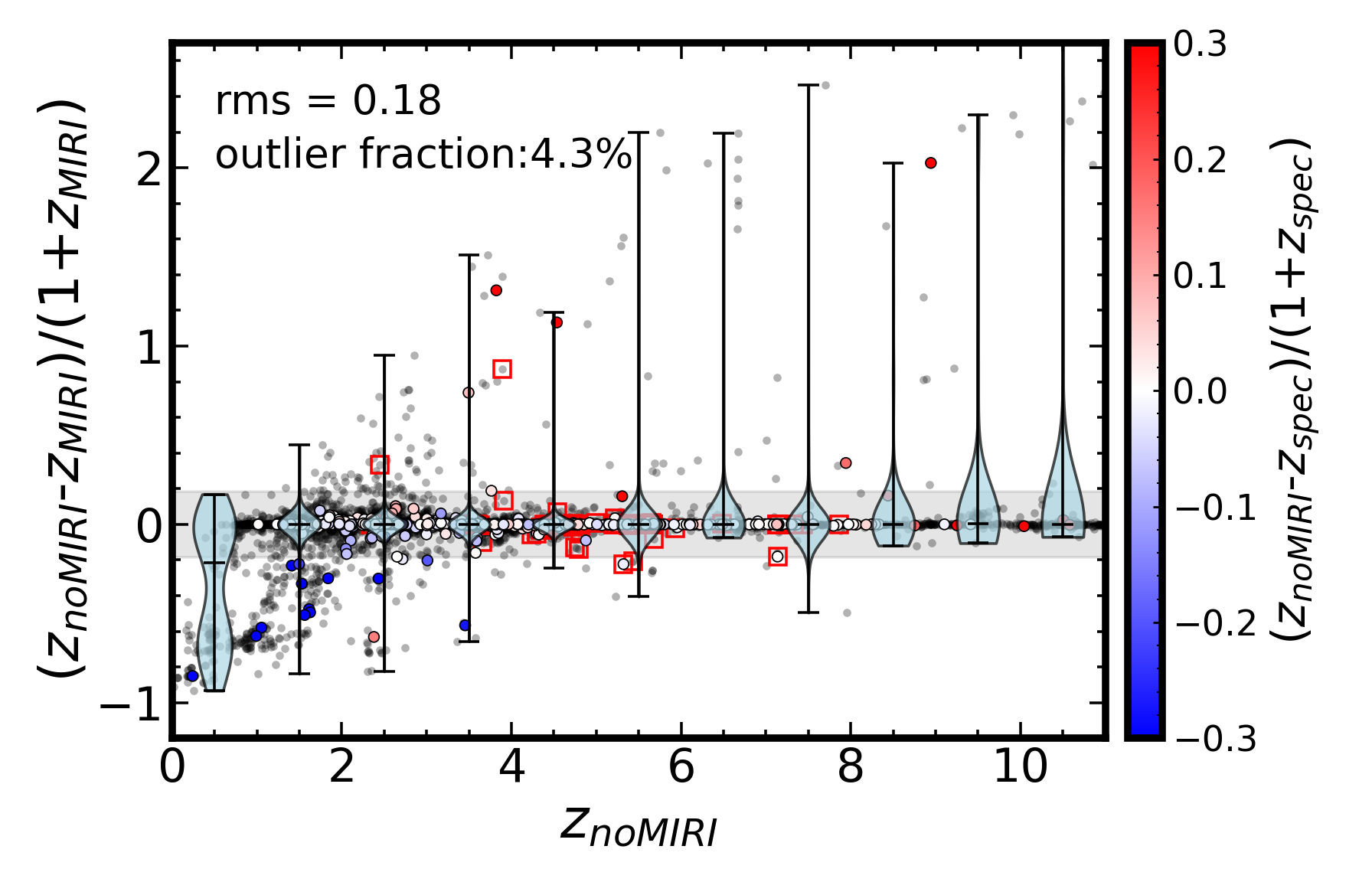} 
	\caption{\small{\textbf{Impact of MIRI on photometric redshift estimates.} The data points indicate the relative difference between the derived photo$-z$s for the galaxies excluding MIRI ($z_{\rm noMIRI}$) and when including MIRI ($z_{\rm MIRI}$, $\Delta z = (z_{\rm no MIRI} - z_{\rm MIRI})/(1 + z_{\rm MIRI}$)) as a function of $z_{\rm noMIRI}$. The rms values and outlier fractions are indicated in the top-left corner. Galaxies with spec-zs are further color-coded by the relative differences between spec-zs and $z_{\rm noMIRI}$, $\Delta z_{\rm spec} = (z_{\rm no MIRI} - z_{\rm spec})/(1 + z_{\rm spec}$)). The red filled circles and open squares show LRDs, with the open squares indicating sources with spec-zs. The violin plots denote the median and 1$\sigma$ confidence interval of $\Delta z$ at each redshift bin.
}}
	\label{efig:z_comp}
\end{figure*}

We evaluate here the impact of adding MIRI photometry on the derivation of photo-zs. Figure~\ref{efig:z_comp} shows the comparison between photo-zs derived with and without MIRI photometry.  Although generally no systematic differences are found for the majority of the sample with a median differences close to zero, there is a trend that many galaxies with $z_{\rm noMIRI} \lesssim 4$ are actually at higher $z$ when fitted with MIRI, while this trend reverses at $z_{\rm noMIRI} \gtrsim 4$.
Similar trend is found when compared $z_{\rm noMIRI}$ to spec-zs, which proves that better photo-zs can be derived after including MIRI. 

\begin{figure*}
	\centering
	\includegraphics[scale=0.42]{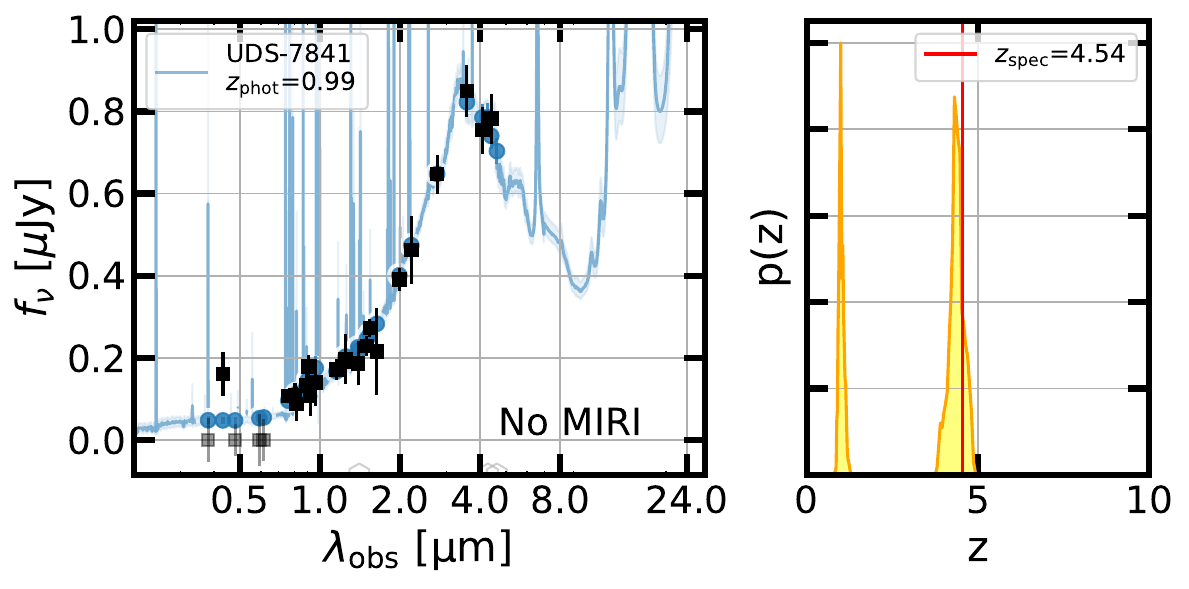}
    \includegraphics[scale=0.42]{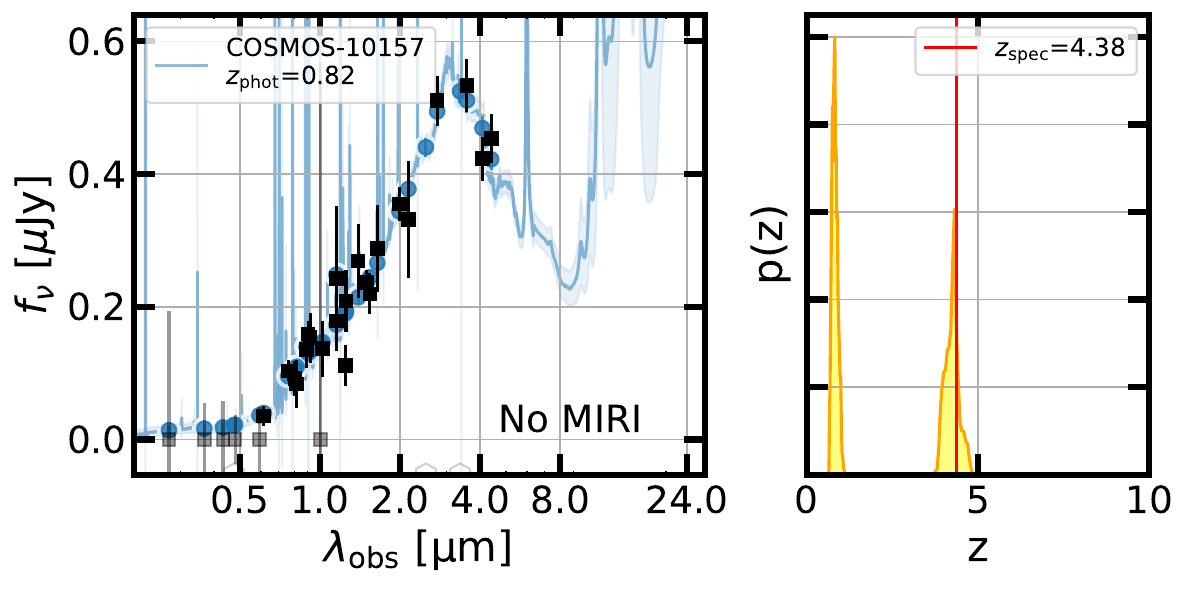}
    \includegraphics[scale=0.42]{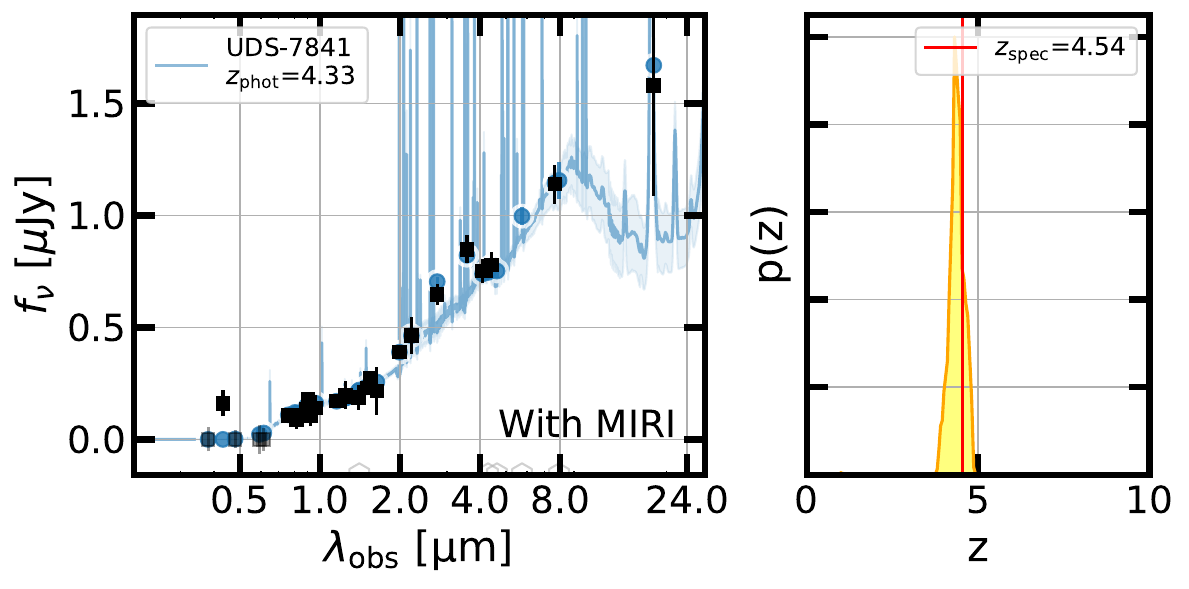}
    \includegraphics[scale=0.42]{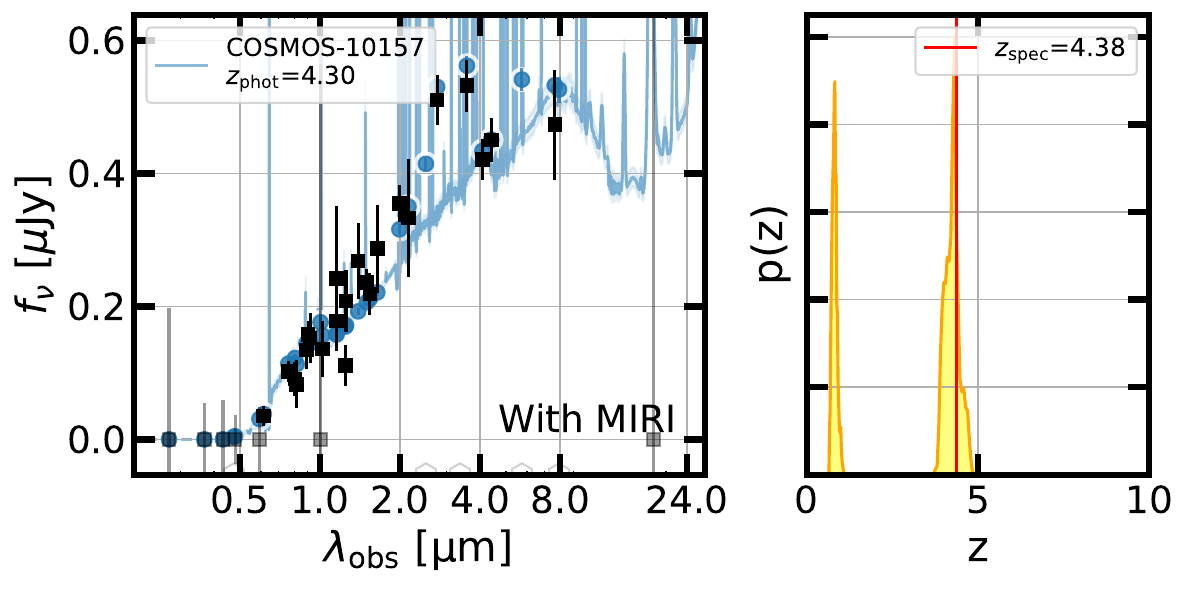}
	\caption{\small{\textbf{Examples SEDs and photo$-z$ PDFs for galaxies with underestimated $z$ without MIRI.}  The top two panels show the best-fit SEDs and PDFs of their photo-zs when fitted without MIRI, while the bottom two panels show results when MIRI is included. The spec-z of each source is indicated by the red vertical line with their values shown in the top-right corner of the p($z$) plots.
}}
	\label{efig:z_comp_example}
\end{figure*}

For galaxies with systematic lower redshifts when fitted without MIRI at $z_{\rm noMIRI} \lesssim 4$, we find that they often exhibit multiple peaks in the probability distribution function (PDF) of their photo-zs. This is mainly caused by the difficulty in distinguishing between optical emission lines and the 1.6$\mu$m stellar bump when MIRI is not included. The addition of MIRI, including both F770W and F1800W, suppresses the probability of the low-z solution. We show two such cases in Figure~\ref{efig:z_comp_example}.

For galaxies with higher photo-zs when fitted without MIRI at $z_{\rm noMIRI} \gtrsim 4$, the limited samples with spec-zs do not allow for a clear answer to their nature. some of them are likely AGNs with strong emission lines, for which both $z_{\rm noMIRI}$ and $z_{\rm MIRI}$ turn to deviate significantly from spec-zs although in most cases $z_{\rm MIRI}$ tend to underestimate spec-zs, unlike the case for $z_{\rm noMIRI}$. Future studies with much larger spectroscopic samples will be needed to fully address these issues.

\begin{figure*}
	\centering
	\includegraphics[scale=0.18]{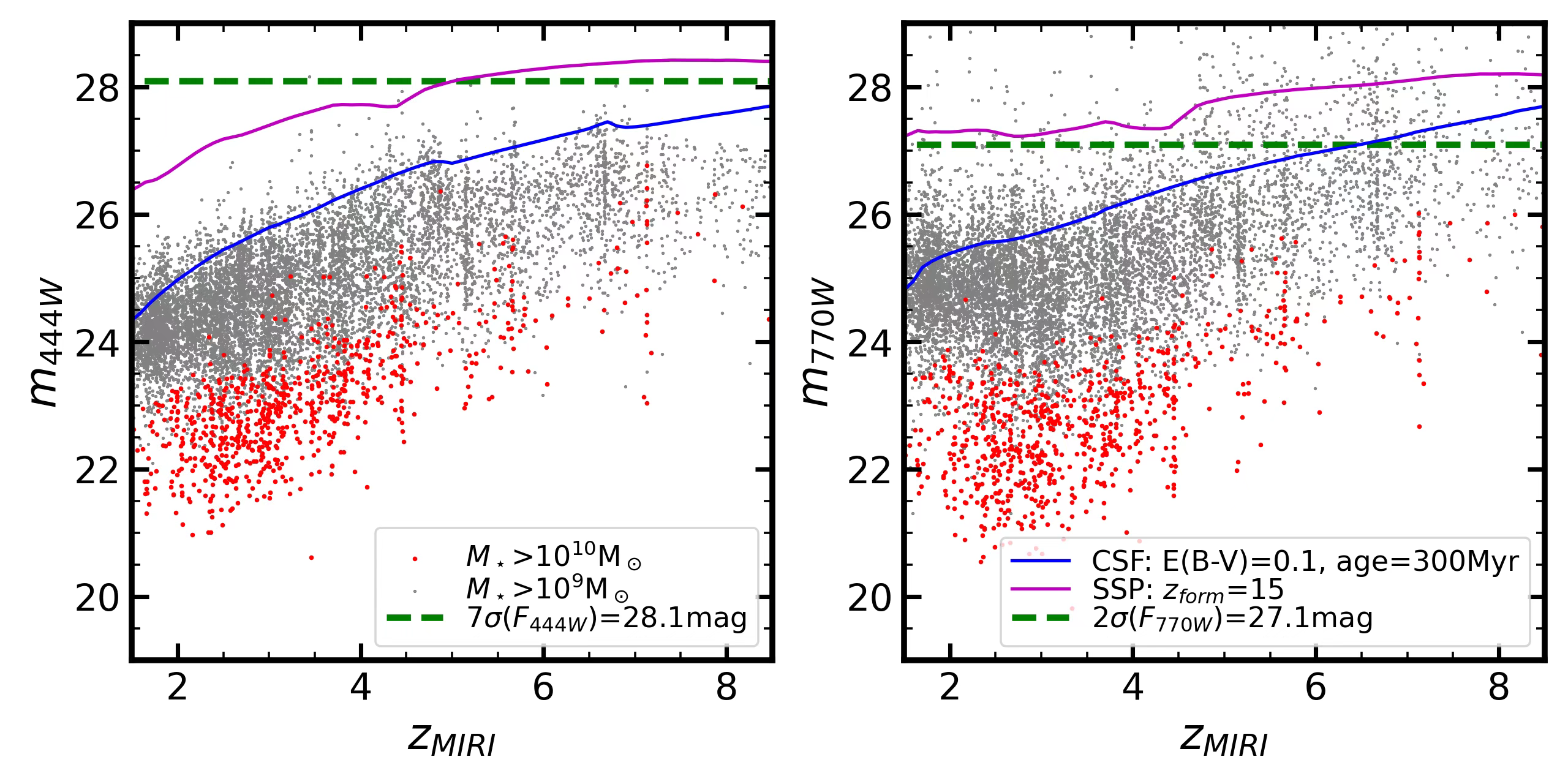} 
	\caption{\small{\textbf{Mass completeness vs. redshift at different magnitudes for our galaxy sample.} The left panel shows the MIRI/F770W magnitudes while the right panel shows the NIRCam/F444W magnitudes. The green dashed lines show detection limits (7$\sigma$ for F444W, 2$\sigma$ for MIRI/F770W). The purple solid lines indicate the corresponding magnitudes for an old galaxy template formed at $z = 15$ at $\mstar = 10^{9} M_{\odot}$. The same mass limits are shown with the blue solid line for a 300 Myr SFG template with constant star formation (CSF) history and $E(B-V)$ = 0.1. The black and red data points denote respectively the galaxies with $\mstar > 10^{9} M_{\odot}$ and $\mstar > 10^{10} M_{\odot}$. 
}}
	\label{efig:mass_completeness}
\end{figure*}

\section{mass completeness}
\label{appendix:mass_complete}
We derive a conservative mass completeness using a maximum old galaxy template formed at $z = 15$ through a single burst. Figure~\ref{efig:mass_completeness} shows the corresponding F444W and F770W magnitudes of this old galaxy template at different redshfits. We also show a 300-Myr star-forming galaxy template with a constant star formation history, which is found to better match the actual data.  The 7$\sigma$ detection limit at F444W is 28.1 mag for point sources, ensuring that all massive galaxies with $\mstar > 10^{9} \msun$ are included in our sample up to $z = 8.5$. In addition, the right panel shows that while not every galaxy with $\mstar > 10^{9} \msun$ is significantly detected at F770W, all the most massive ones with $\mstar > 10^{10} \msun$ are detected at least at the 2$\sigma$ level up to $z = 8.5$.

\begin{figure*}
	\centering
	\includegraphics[scale=0.4]{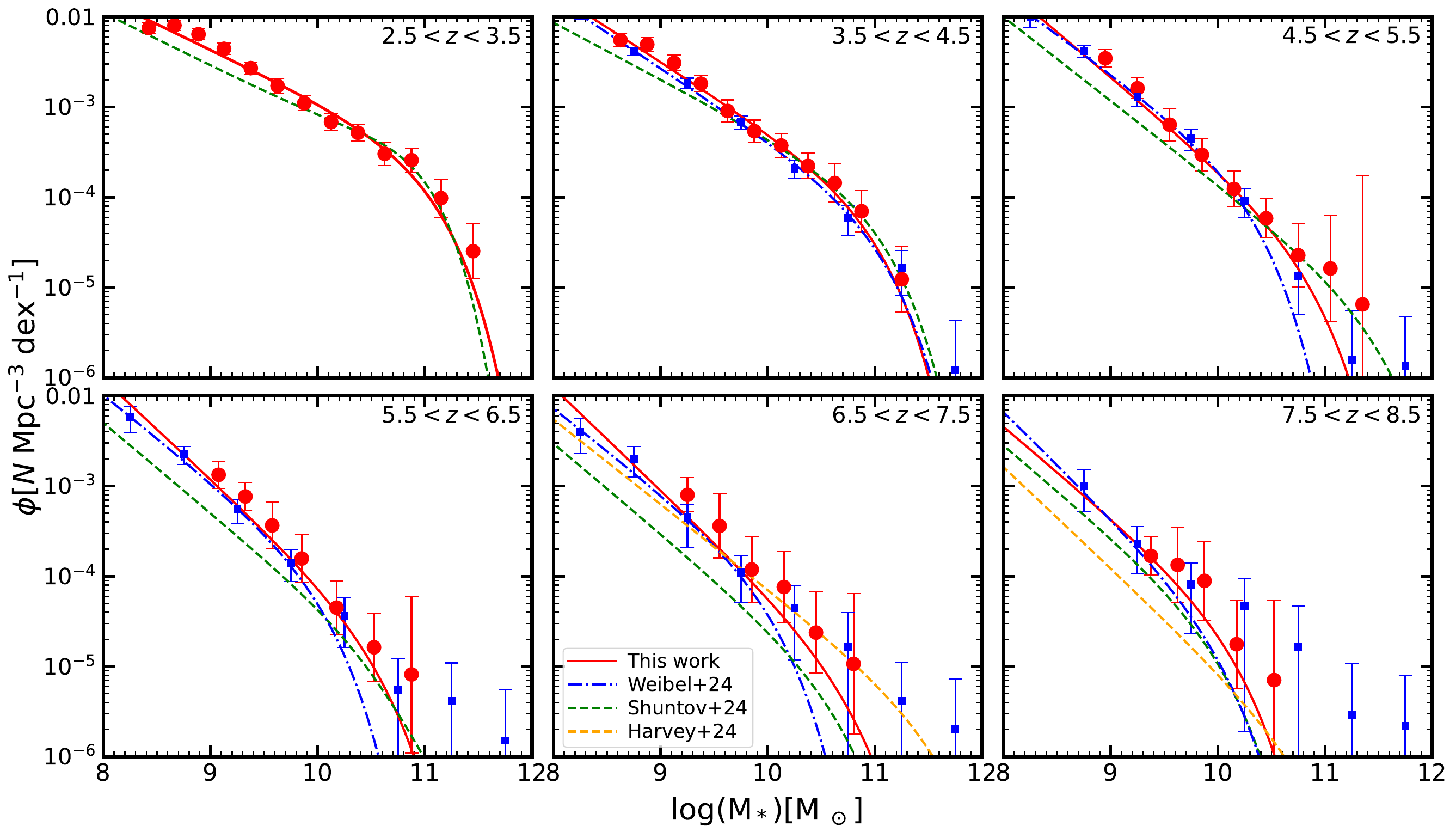} 
	\caption{\small{\textbf{Comparison with previous SMF studies based on mainly NIRCam data .} The filled red circles denote the observed values based on our primary sample (i.e., excluding LRDs) with MIRI constraints. The red solid lines indicate the best-fit Schechter function (before convolution) of this work, while the other lines show the SMFs from other recent works based on mainly JWST/NIRCam data \citep{Harvey2024,Shuntov2024,Weibel2024}.
}}
	\label{efig:SMF}
\end{figure*}
\section{Comparison with recent SMFs studies}
\label{appendix:SMF_comp}
Figure~\ref{efig:SMF} compares our SMFs at $z \sim 3-8$ with previous {\it JWST} studies, which mainly used NIRCam data. Our results show excellent agreement at $z \sim 3-5$ over all mass ranges with ~\cite{Weibel2024}, which combined the public JWST/NIRCam imaging programs CEERS, PRIMER, and JADES. At $z > 5$, our low-mass end still agrees, but the massive end is lower. This is consistent with MIRI's effects on $\mstar$. Compared to the recent estimate of SMFs from COSMOS-Web~\citep{Shuntov2024}, while our derived SMFs (and those of ~\cite{Weibel2024}) agree at $z \sim 3-4$, significant differences are found at $z > 5$. Given the consistent results between ours, ~\cite{Weibel2024}, and also with ~\cite{Harvey2024} at least at $z \sim 7$, we argue that this difference is unlikely due to cosmic variance. Rather, given that COSMOS-Web includes only four JWST/NIRCam bands, we argue that this difference may be due to the relatively poor photo-z quality at $z \gtrsim 5$ for COSMOS-Web galaxies. A detailed investigation on the origin of different studies is beyond the scope of this paper, which we leave for future work.


\bibliographystyle{aasjournal}
\bibliography{ref.bib}
\end{document}

%% file: extra_preamble.tex
\newcommand{\mstar}{M_{\star}}
\newcommand{\msun}{M_{\odot}}

\newcommand{\lcdm}{$\Lambda$CDM}

\newcommand{\muv}{{M_{\rm UV}}}

\newcommand{\fbary}{f_{\rm b}}